\documentclass[11pt] {article}      
\usepackage{amsmath, amssymb, amsthm, eucal, accents}  
\usepackage{graphicx}
\usepackage{txfonts}
\usepackage[T1]{fontenc}     %for better hyphenation etc
\usepackage{enumerate}       %used for better lists
\usepackage{float}                 % use in \begin{figure}[H] to enforce position
\usepackage{mathtools}
\usepackage{tikz}
\usetikzlibrary{matrix, arrows, calc, fit}
\usetikzlibrary{decorations.markings}      %}
\tikzstyle{AArrow} = [thick, decoration={markings,mark=at position 1 with {\arrow[semithick]{open triangle 60}}},% 
   double distance=1.4pt, shorten >= 5.5pt,
   preaction = {decorate},
   postaction = {draw,line width=1.4pt, white,shorten >= 4.5pt}]
\tikzstyle{AArroww} = [semithick, white,line width=1.4pt, shorten >= 4.5pt]
\usepackage[strict]{changepage}
\usepackage{relsize}
\theoremstyle{plain}
\newtheorem{theorem}{Theorem}[section]            %{commend}{in print}[numbered within]       
\newtheorem{proposition}[theorem]{Proposition}  %{commend}[numbered with]{in print as}
\newtheorem{lemma}[theorem]{Lemma}
\newtheorem{corollary}[theorem]{Corollary}	      %	      

\theoremstyle{definition}
\newtheorem{definition}[theorem]{Definition}

\numberwithin{theorem}{section}
\numberwithin{equation}{section}
\numberwithin{figure}{section}
\newcommand{\gaction}[2]{\genfrac{}{}{0.5pt}{}{#1}{#2}% 
                        \!\lower2pt\hbox{\rotatebox[origin=c]{-90}{{$\looparrowright$}}}}
\newcommand{\dgaction}{\displaystyle\gaction}
\newcommand{\dotfraction}[2]{\genfrac{}{}{0.5pt}{}{#1}{#2}% 
                        \!\lower.5pt\hbox{{$\circ$}}}
\def\mathbi{\boldsymbol}   		%good for italic bold i, for instance

\def\rmspan{\hbox{span}\,}

\def\rmIm{\hbox {\rm Im\,}}
\def\rmspan{\hbox {\rm span}\,}
\def\q{\partial}

\renewcommand\appendix{\par
  \setcounter{section}{0}
  \setcounter{subsection}{0}
  \setcounter{figure}{0}
  \setcounter{table}{0}
  \renewcommand\thesection{Appendix \Alph{section}}
  \renewcommand\thefigure{\Alph{section}\arabic{figure}}
  \renewcommand\thetable{\Alph{section}\arabic{table}}
}
\usepackage{titlesec}                                                           %fixes section titles
\titlelabel{\thetitle.\ }                                                           %number
\titleformat*{\section}{\fontsize{14pt}{14pt} \bf}                   %size
\usepackage{fancyhdr}
\usepackage{sidecap}  %for side caption of figures.  Use :
\usepackage[hang,small,sc]{caption}

\def\smalll{\scriptsize}

\def\QED{ $\square$}

\def\ee{\varepsilon}
\def\ff{\varphi}

\def\moplus{\boxplus}
\def\voplus{\oplus}

\def\rmId{\hbox{\rm Id}}
\def\rmgen{\hbox{gen}\;}
\def\Rev{\hbox{\rm Rev}}
\def\rmid{\hbox{\rm id}}
\def\rmid{\hbox{\rm id}}
\def\Inv{\hbox{\rm Inv}}
\def\Mat{\hbox{\rm Mat}}

\begin{document}

\title{\bf %A new geometric representation of the Lorentz group and the relativistic composition of velocities %
%Cromlech, menhirs and celestial sphere: a new approach to relativistic composition of velocities % relativistic composition of velocities
%Lorentz group, celestial sphere and relativistic composition of velocities: algebra and geometry % relativistic composition of velocities
%Lorentz group, celestial sphere, cromlech  and relativistic composition of velocities
%Lorentz group via reversions: celestial sphere and relativistic composition of velocities
Cromlech, menhirs and celestial sphere: an unusual representation of the Lorentz group
% Relativistic Stonehenge, All done with a meadow and stones only
}
\author{Jerzy Kocik                  %\thanks{support}
\\ \small Department of Mathematics
\\ \small Southern Illinois University, Carbondale, IL62901
\\ \small jkocik{@}siu.edu  }
%\date{\small (Sep 2011)}
\date{} %{\small -- Version 12 April 2016 --}

\maketitle

\begin{abstract}
\noindent
\small 
We present  a novel representation of the Lorentz group, the geometric version of which 
uses ``reversions'' of a sphere while the algebraic version uses pseudo-unitary $2\times 2$ matrices     
over complex numbers and quaternions, and Clifford algebras in general.
A remarkably simple formula for relativistic composition of velocities
and an accompanying geometric construction follow.
The method is derived from the diffeomorphisms of the celestial sphere induced by Lorentz boost. 
%We present a remarkably simple formula for relativistic composition of velocities
%and an accompanying geometric construction.
%It is based a novel representation of the Lorentz group, the geometric representation of which 
%uses ``reversions'' of a sphere, and the algebraic version uses    
%Clifford algebras, starting with the complex numbers and quaternions for small dimensions.
%The method is derived from the diffeomorphisms of the celestial sphere induced by Lorentz boost. 
\\[3pt]
%\smalskip\noindent
{\bf Keywords:}   $\hbox{\rm SU}(1,1)$, inversive geometry, reversions, complex numbers, quaternions, Clifford algebra,
relativity, space-time, composition of velocities, Thomas rotation, celestial sphere, golden ratio, 
visualization, 
%Stonehenge, 
menhir.%
\\[5pt]
\noindent \textbf{AMS Subject classification}: 
83A05, %~(Spec. rel.), 
15A23, %~(Factorization of matr's), 
15A66, %~~(Clifford alg, spinors), 
15A90, %~(Appl. of matrix th. to phys), 
51P05. %~(Geo and phys.)
\\
\noindent \textbf{PACS}:
03.30.+p, 	% Special relativity
02.40.Dr,  	% Euclidean and projective geometries
02.10.Yn. %	Matrix theory
\end{abstract}

%%%%%%%%%%%%%%%%%%%%%%

%-------------------------------------------------------------------------------------------------------------------------------
\section{ Introduction and results}

\begin{figure}[h]
\centering
\includegraphics[scale=.7]{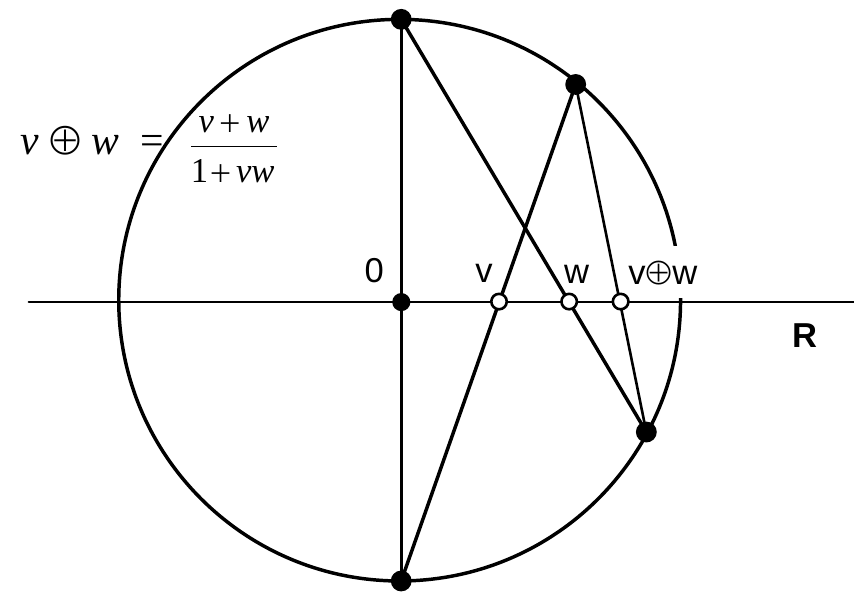}
\caption{\small Visualization of the Poincar\'e formula \eqref{eq:intro1} for velocities
as points \textit{v} and \textit{w} on real line $\mathbb{R}$.
The velocity of relativistic composition $v\voplus w$
is determined by the quadrilateral inscribed in the unit circle as shown.}
\label{fig:vdiagram}
\end{figure}

In one dimension, the formula for relativistic composition of velocities has a simple algebraic form:
%$$
\begin{equation}
\label{eq:intro1}
v \voplus w =   \frac{v+w}{1+vw} \,,
\end{equation}
%$$
discovered in 1904 by Henri Poincar\'e \cite{Po}.  
The above operation turns the open interval $(-1,1)$ into an Abelian group.  %
%\footnote{The formula for hyperbolic tangent angle addition in disguise.}
It may be visualized in terms of a simple geometric gadget 
reproduced in Figure \ref{fig:vdiagram} from \cite{jk-diagram, jk-porism}.
Once we go beyond collinearity of velocities, the algebraic simplicity of the Poincar\'e formula is lost. 
%\cite{Ri}.  
This is because the boosts, which are among the generators of the Lorentz Lie group ${\rm SO}(1,n)$,  
do not form a subgroup and a rotational component emerges in a product of two boosts. 
%Much of literature is devoted to this ``nuisance''; see e.g.,  

%There are attempts to give the notion of ``velocity adding'' some geometric simple picture 
%mostly based on hyperbolic geometry of Poincar\'e disk \cite{RS, Wi}.  

%================= FIGURE 1

\begin{figure}[ht]
\begin{minipage}[b]{0.5\linewidth}
     \centering
     \includegraphics[scale=.74]{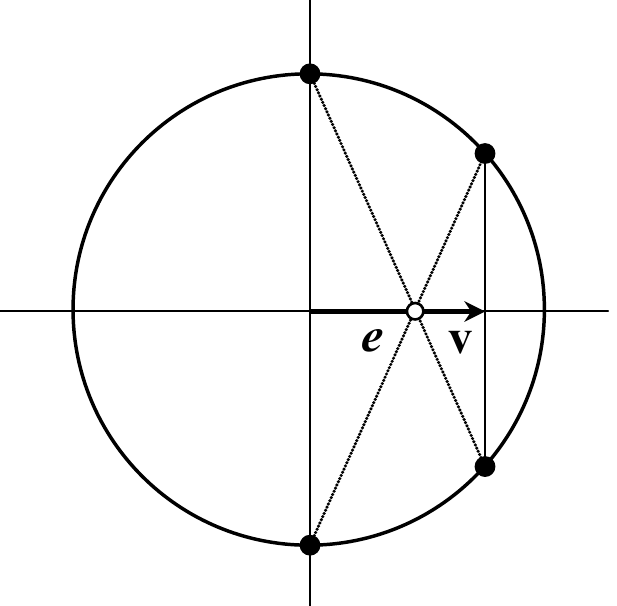}
     \caption{\small How to find the menhir for a given velocity.}
     \label{fig:intro-ev}
\end{minipage}
\hspace{0.5cm}
\begin{minipage}[b]{0.5\linewidth}
      \centering
     \includegraphics[scale=.71]{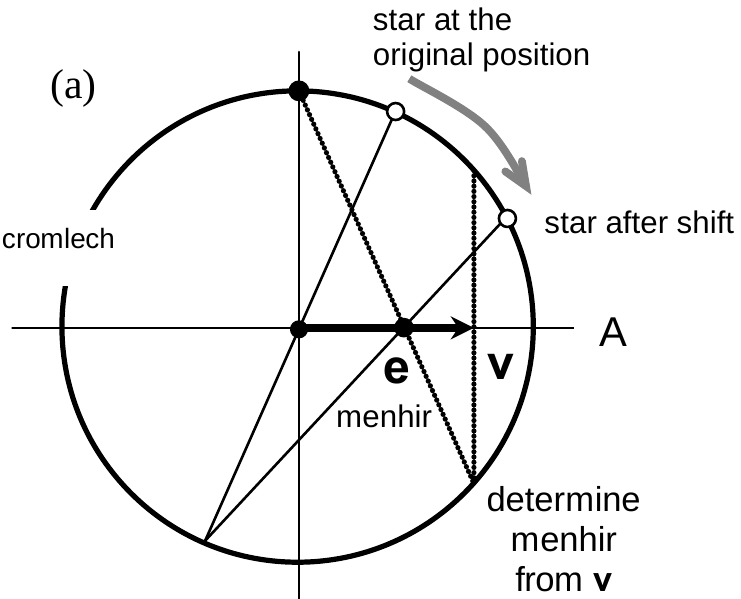}
     \caption{\small %First,  determine the location of  the menhir or the obelisk (dotted lines) for a given velocity.  
          Trace the shift of any star due to boost using the menhir.}
     \label{fig:intro3}
\end{minipage}
\end{figure}

In the present article, we propose an approach which has two expressions: algebraic and geometric.  
The basic idea is to present velocities in a reduced form called \textit{menhirs},  
the construction of which is shown in Figure \ref{fig:intro-ev}.
%They constitute -- quite like the regular velocities -- a unit disk in space.  
The two formalisms are:
%\newpage

\begin{itemize}
\item
{\bf Geometric representation}  --- composition of boosts obtains a very simple %and elegant 
form in terms of geometric constructions based on inversions (``reversions'') of a sphere
through the points called menhirs.  \\[-17pt]
\item
{\bf Algebraic representation}  --- In the algebraic form, we use division algebras like $\mathbb R$, $\mathbb C$ or $\mathbb H$ 
and other Clifford algebras.
Remarkably, the  ``menhir representation'' of velocities restores the Poincar\'e-like simplicity of the formula for the composition of boosts.
%With the use of complex numbers or quaternions, the calculations become a breeze.  
\end{itemize}

%In both forms,  the Thomas rotation appears naturally and the relativistic deformation of the celestial sphere can be easily traced.  
%
%\noindent \textbf{A.  Graphical composition of velocities.}  
%Two collinear velocities are presented as points \textit{v} and \textit{w} on real line $\mathbb{R}$, see Figure \ref{fig:vdiagram}.  
%Add to the figure a unit circle centered at the origin $0$.  
%Draw two lines, one through $(0,1)$ and $v$ and the other through $(0,-1)$ through $w$.
%Their points of intersection with the circle, when joined, determine the point on $\mathbb R$ that coincides  with
%the velocity of relativistic composition $v\voplus w$ of Equation \eqref{eq:intro1}.   

\begin{figure}[h]
\begin{tikzpicture}[baseline=-0.8ex]
    \matrix (m) [ matrix of math nodes,
                         row sep=2.5em,
                         column sep=3.2em,
                         text height=4ex, text depth=3ex] 
 {
           \quad \dgaction{\hbox{\rm SO}_{\rm o}(1,n)}{\mathbb R^{1,n}}   \quad
          & \  % \quad \dgaction{SU(2)}{P\mathbb C^2}  \quad   
          &\quad  \dgaction{\hbox{\rm SU}_{1,1}(\, \mathbb F\, )}{~ \mathbb F^{\,2} ~}  \quad    \\
           \quad \dgaction{\hbox{\rm SO}_{\rm o}(1,n)}{\mathbb R^{1,n}_{0}}   \quad 
          & \quad \dgaction{\hbox{gen}\,\{\mathbf p|p\in D^n\}}{S^{n-1}=\q D}   \quad   
          &\quad  \dgaction{\hbox{\rm PSU}_{1,1}(\,\mathbb F\,)}{{\mathbb F\cup\{}\infty\}}  \quad     \\
           \quad \dgaction{\hbox{\rm PSO}_{\rm o}(1,n)}{~\hbox{\rm P}\mathbb R^{1,n}_{0}\cong S^{n-1}~}   \quad 
          & \quad \dgaction{\Rev_{\rm o}(n)}{K\cong S^{n-1}}   \quad   
          &\quad  \dgaction{\hbox{\rm PSU}_{1,1}(\,\mathbb F\,)}{S^{n-1}\subset  \mathbb F ~  }  \quad     \\
  };
            \path[triangle 60-]        (m-1-1) edge node[above] {\small $2:1$}  (m-1-3);
            \path[triangle 60-triangle 60]   (m-3-1) edge node[above] {\small $1:1$} node [below] {$\cong$}   (m-3-2);
            \path[triangle 60-triangle 60]        (m-3-2) edge node[above] {\small $1:1$} node [below] {$\cong$}  (m-3-3);
%vertical
    \path[-triangle 60]        (m-1-1) edge node[right] {$\subset$}  (m-2-1);
    \path[-triangle 60]        (m-1-3) edge node[right] {$\hbox{proj}$}  (m-2-3);
    \path[-triangle 60]        (m-2-1) edge node[right] {$\hbox{proj}$}  (m-3-1);
    \path[triangle 60-]        (m-2-2) edge node[right] {$\subset$}  (m-3-2);
    \path[-triangle 60]        (m-2-3) edge node[right] {$\subset$}  (m-3-3);

\node at (m-1-3.north) [above=-1pt, color=black] {\smalll\sf {M\"obius action}};
\node at (m-2-2.north) [above=1pt, color=black] {\smalll\sf {reversions}};
\node at (m-1-1.north) [above=-1pt, color=black] {\smalll\sf {relativity}};

\node at (m-1-3.north) [above=7pt, color=black] {\smalll\sf {ALGEBRA}};
\node at (m-2-2.north) [above=9pt, color=black] {\smalll\sf {GEOMETRY}};
\node at (m-1-1.north) [above=7pt, color=black] {\smalll\sf {PHYSICS}};

\node at (m-3-1.south) [below=7pt, color=black] {\smalll\sf {CELESTIAL}};
\node at (m-3-2.south) [below=7pt, color=black] {\smalll\sf {CROMLECH}};
\node at (m-3-3.south) [below=7pt, color=black] {\smalll\sf {matrices}};
\node at (m-3-1.south) [below=15pt, color=black] {\smalll\sf {SPHERE}};

\end{tikzpicture}   
\caption{The diagram of the main points, the 2+1 dimensional Minkowski space case}
\label{fig:bigdiagram}
\end{figure}
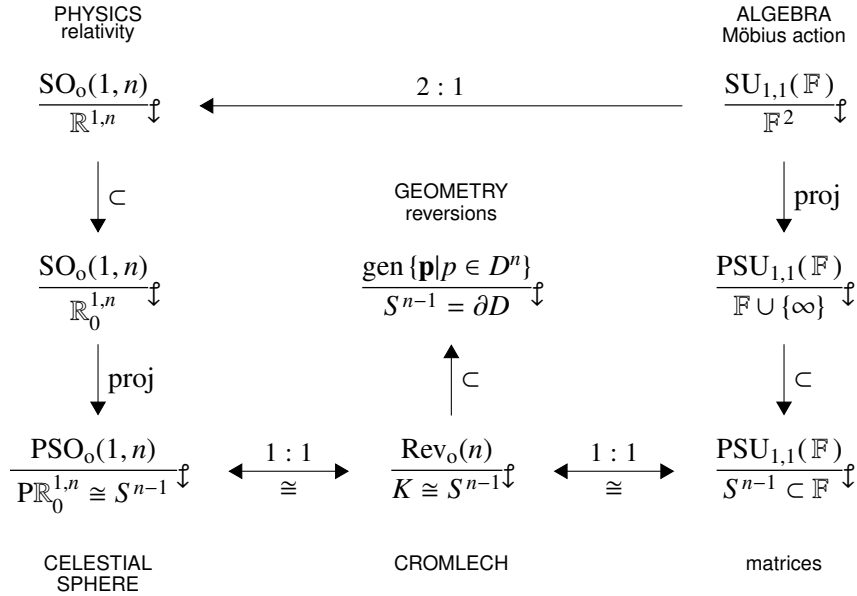

~\\
{\bf Summary:} %of the ``practical'' content of the .... for the clarificationd 
We present the material as a ``tail of three groups''
by taking a trip through the diagram  of 
homomorphisms and relations presented in Figure \ref{fig:bigdiagram}.
%The commutative between celestial action of the Lorentz group, geometric reversions of a sphere,    
The fraction-like symbol $\gaction{G}{X}$ indicates that the group $G$ acts on the set $X$.
The three columns are the three wings of the construction: (1) relativity and Lorentz group,
(2) the geometry of sphere, 
and (3) its algebraic version in terms of certain matrix groups.
Here is the general  legend to this diagram.

\begin{enumerate}

%\item
%The fraction-like symbol $\gaction{G}{X}$ means that the group $G$ acts on the set $X$.
%\\[-17pt]
%\item
%The three columns are the three wings of the construction: (1) relativity with Lorenz group (2) the geometry of sphere, 
%and (3) its algebraic version.
%\\[-17pt]
\item 
The left top corner denotes the Lorentz group acting on the Minkowski vector  space $M\cong \mathbb R^{1,n}$.
This action may be restricted to the isotropic vectors, the light cone $\mathbb R^{1,n}_0$ (map ``$\subset$'').
This induces  an action on {\it rays} in the light cone, elements of the subset ${\rm P}\mathbb R^{1,n}_0$ of the projective  space.
Topologically it is a sphere, ${\rm P}\mathbb R^{1,n}_0\cong S^{n-1}$, called the {\bf celestial sphere}.
The Lorentz group can be perceived via  deformations of the celestial sphere, the {\bf aberration} of its points.
\\[-17pt]
\item
An observer understood as a space-like subspace of $\mathbb R^{1,n}$ may identify   
the celestial sphere with the unit sphere $S^{n-1}$  (called here {\bf  cromlech}) in her space.
We show that every velocity may be represented by a point $p\in D\subset \mathbb R^n$ called {\bf menhir} inside the cromlech
via a certain ``menhir map'' and show how this point determines the aberration in a simple purely geometric way. 
\\[-17pt]
\item
The last column refers to the matrix representation of the reversions sphere
with the use of a field $\mathbb F$.  
In the case of (1+2)-Minkowski space $n=2$,  $\mathbb F=\mathbb C$ and the group is the unitary group $\hbox{\rm SU}(1,1)$
acting on $S^1\subset \mathbb C$ via the fractional linear maps.
The case of $n=4$ involves quaternions, $\mathbb F=\mathbb H$, 
the corresponding group is $\hbox{\rm SU}_{1,1}(\mathbb H) \equiv {\rm Sp}(1,1)$, and the celestial sphere is $S^{3}$.
%(Interestingly, these groups coincide with the spin groups of the  Lorentz group.  
%(For more on $n=2$ case, see \cite{jk-clifford}).
This wing leads to a simple algebraic formula for addition of velocities.
A generalization to other Clifford algebras takes care of $n>4$.
\end{enumerate}

~

%---------------------------------------------------------

%-----------------------------------------

\noindent 
\textbf{A.  Menhirs -- a representation of velocities.}  We shall associate with every velocity $\mathbf v$ 
a point (vector) 
%-- ``stone marker'' will be 
$\mathbf e = \mu(\mathbf v)$, called the \textbf{menhir} (this megalithic lingua will make sense soon)
defined by construction shown in Figure \ref{fig:intro-ev}. 
The resulting algebraic relation between the two is:
%$$
\begin{equation}
\label{eq:intro2}
\mathbf v=\frac{2\mathbf e}{1+e^{2} } 
\end{equation}
%$$
One may think of $\mathbf e$ as a non-uniformly {\it rescaled} velocity $\mathbf v$:
%Although the rescaling is not uniform, still $\mathbf v = 0$ corresponds to $\mathbf e = 0$,  and speed of light $v = 1$ corresponds to $e = 1$.
%Despite this distortion -- the relation between the ``menhirs'' corresponding to $v_1$, $v_2$ 
%and $v_1\voplus v_2$ of relativistic addition \eqref{eq:intro1} of collinear velocities satisfy the same rule:
%%$$
%\begin{equation}
%\label{eq:intro4}
%e = e_1\moplus e_2   =  \frac{e_{1} +e_{2} }{1+e_{1} e_{2} } 
%\end{equation}
%%$$
%\noindent \textbf{With the menhir we can trace celestial sphere deformations due to boosts.}  

%
%%================= FIGURE 1
%\begin{figure}[h]
%\centering
%\includegraphics[scale=.71]{C-intro1}
%\caption{\small How to find the obelisk and the menhir.}
%\label{fig:intro2}
%\end{figure}
%
%========= Fig 

With the menhirs we can trace celestial sphere deformations due to boosts.  
Let us present it in the two-dimensional case.  Stars on the horizon form a visual apparent circle.  
We may describe their celestial positions by points (stones) on a unit circle $K$ that we will call \textbf{cromlech}.  
Under a boost the stars  change their apparent positions: 
they will all be shifted towards the front-most point, say, \textit{A} on the cromlech, 
except the star at $A$ and its antipodal star, $-A$, which stay invariant (see Figure  \ref{fig:intro3}).
We will show that one can easily ``predict'' the shift of each star using the menhir.
Simply, reflect the ``star'' through the center and then through the menhir.  
%Such a ``two-stone'' algebra handles more general transformations of the Lorenz group $SO(1,n)$.
%
%
%%================= FIGURE 1
%\begin{figure}[H]
%\centering
%\includegraphics[scale=.71]{C-intro2}
%\caption{\small First,  determine the location of  the menhir or the obelisk (dotted lines) for a given velocity.  
%Then you may trace the shift of any star due to boost.}
%\label{fig:intro3}
%\end{figure}
%
%
\\
\\
\noindent \textbf{B.  Algebraic version of menhir calculus.}  
The geometry of menhirs can  be given algebraic form  in terms of division algebras $\mathbb F = \mathbb R, \mathbb C, \mathbb H$ 
(real numbers, complex numbers and quaternions) which take care of 1-, 2-, and 4-dimensional space. 
Both the velocity and its menhir become elements of ~$\mathbb F$ inside the unit disk $D\subset\mathbb F$.
%circle: $D^1\subset \mathbb R$, $D^2\subset \mathbb C$, or $D^4\subset \mathbb{H}$.  
We shall use Greek letters when the menhirs are understood as algebra elements, i.e., , 
$\varepsilon\in\mathbb F$ for $\mathbf e$.  
Now, the map 
$$
\mu: D \to D: \ v\mapsto \varepsilon = \mu(v)
$$
defined geometrically in Figure \ref{fig:intro-ev} translates into relations:
%$$
\begin{equation}
\label{eq:intro10}
 v  
%=  \frac{2\varepsilon }{1+\bar{\varepsilon }\varepsilon }  
= \frac{2\varepsilon }{1+|\varepsilon |^{2} } \,,
\qquad
\varepsilon  =  \frac{v}{1+\sqrt{1-|v|^2} } \,,
\end{equation}
%$$
%(corresponding to geometry in Figure \ref{fig:intro-ev}). 
where $|\varepsilon |^2= \varepsilon \bar\varepsilon$.
 A composition of two boosts, described by menhirs $\varepsilon_1$ and $\varepsilon_2$,  
is equivalent to a single boost described by menhir 
%$$
\begin{equation}
\label{eq:intro11}
\varepsilon_1 \moplus \varepsilon_2   =  \frac{\varepsilon _{1} +\varepsilon _{2} }{1+\bar{\varepsilon }_{1} \varepsilon _{2} } 
\end{equation}
%$$
(which---quite remarkably---has the same form as the Poincar\'e formula \eqref{eq:intro1}), 
followed by a (Thomas) rotation represented by 
%$$
\begin{equation}
\label{eq:intro12}
\rho =  \frac{1+\varepsilon _{2} \bar{\varepsilon }_{1} }
                  {1+\bar{\varepsilon }_{2}\varepsilon _{1}  } 
\end{equation}
%$$
%Note that \eqref{eq:intro11} boils down to \eqref{eq:intro4} for collinear velocities, or the case of $\mathbb F=\mathbb R$. 
%But in general the product \eqref{eq:intro11} it is not commutative and does not turn the disk $D^2$ into a group.  
%\\

\noindent 
These two equations may be read off from a single matrix ``master equation''
%$$
\begin{equation}
\label{eq:intro17}
\left[\begin{array}{cc} {1} & {\varepsilon _{2} } \\ {\bar{\varepsilon }_{2} } & {1} \end{array}\right] \; 
\left[\begin{array}{cc} {1} & {\varepsilon _{1} } \\ {\bar{\varepsilon }_{1} } & {1} \end{array}\right]   
=   
\left[\begin{array}{cc} {1+\varepsilon _{2} \bar{\varepsilon }_{1} } & {0} \\ {0} & {1+\bar{\varepsilon }_{2} \varepsilon _{1} } \end{array}\right]\;
\left[\begin{array}{cc} {1} & {\varepsilon _{1} \moplus \varepsilon _{2} } \\ (\varepsilon _{1} \moplus \varepsilon _{2})^*  & {1} \end{array}\right] 
%\left[\begin{array}{cc} {1} & {\varepsilon _{1} \moplus \varepsilon _{2} } \\ {\overline{\varepsilon _{1} \moplus \varepsilon _{2} }} & {1} \end{array}\right] 
\end{equation}
%$$
where the matrices on the left represent the boosts and on the right a boost composed with rotation.  
%This equation is the ``E=mc${}^{2}$'' of the relativistic velocity calculus.
The order of products is here essential due to non-commutativity of quaternions. 
\\

The situation for complex numbers is represented in the following commutative diagram:
%---------------------------------------------------------
\begin{equation}
\begin{tikzpicture}[baseline=-0.8ex]
    \matrix (m) [ matrix of math nodes,
                         row sep=1.7em,
                         column sep=4em,
                         text height=3.2ex, text depth=2.1ex] 
 {
  \hbox{\sf menhirs:} &\quad \mathbb C \times \mathbb C \quad    & \quad \mathbb C  \quad   \\
  \hbox{\sf velocities:} &\quad  \mathbb C\times \mathbb C  \quad     & \quad \mathbb C \quad    \\
  };
    \path[-stealth]
       (m-1-2) edge [transform canvas={yshift=0.5ex}] node[above] {$\moplus$} (m-1-3)
        (m-2-2) edge node[above] {$\voplus$} (m-2-3)
        (m-1-2) edge node[right]  {$\mu\times\mu$} (m-2-2)
        (m-1-3) edge node[right] {$\mu$} (m-2-3)
;
\end{tikzpicture}   
\end{equation}
or,
$
\mu (a \moplus b) = \mu(a)\voplus \mu (b)
$.

~

To clarify, the matrices in \eqref{eq:intro17} form a {\bf group}, namely pseudo-unitary group $\rm SU(1,1)$ 
over one of the division algebras,
homomorphic to the corresponding Lorentz group.
However, operation $\boxplus$ is a non-associative and defines on the unit disk a {\bf loop} $(D,\boxplus)$, 
a quasigroup with an identity.  
The operation of composition of velocities is an isomorphic loop  $(D, \oplus)$.

%============
%The quaternionic fraction is understood as right division:  $a/b = ab^{-1} = a-b / \|b\|^{2}$.  
%In the complex case, $\mathbb{F} = \mathbb{C}$, things get simplified and in particular the term for Thomas rotation becomes  
%%$$
%\begin{equation}
%\label{eq:intro18}
%\rho = e^{i\theta}  =   \frac{1+\bar{\varepsilon }_{1} \varepsilon _{2} }{1+\varepsilon _{1} \bar{\varepsilon }_{2} } 
%=  \frac{(1+\bar{\varepsilon }_{1} \varepsilon _{2} )^{2} }{|1+\bar{\varepsilon }_{1} \varepsilon _{2} |^{2} } 
%\end{equation}
%%$$
%which implies that the angle $\theta$ is twice the Arg$(1+\bar{\varepsilon }_{1} \varepsilon _{2} )$.  
%The quaternionic version is able to deal with five-dimensional space-time $\mathbb R^{1,4}$.  
%For the standard 3-dimensional case one may identify the space with imaginary quaternions 
%$\rmIm\,\mathbb{H} = \rmspan\{\mathbi{i}, \mathbi{j}, \mathbi{k}, \}$.  
%In such a case Thomas rotation -- quite pleasantly -- becomes described by the well-known action of quaternions
%$q\rightarrow pqp^{-1}$ with $p = 1+\varepsilon _{2} \bar{\varepsilon }_{1}$.  
%\\
%
%\noindent In the sections that follow we develop these ideas from the scratch; all concepts will be reintroduced and results proved.
%\\

~
\newpage 

\noindent
{\bf Albebraic conclusion:}  An interesting correspondence emerges between the normed  division algebras and the Lorentz groups:
$$
\begin{array}{clcll|r}
\mathbb R \quad&  \hbox{\rm SO}_{\rm o}(1,1) &\quad \leftrightarrow \quad& \hbox{\rm SU}_{1,1}(\/\mathbb R\/)&\cong \mathbb R 
\qquad &\ \hbox{\rm SO}_{\rm o}(1,2)\ \leftrightarrow \ {\rm SL}(n,\mathbb R) \\
\mathbb C \quad&  \hbox{\rm SO}_{\rm o}(1,2) &\quad \leftrightarrow \quad& \hbox{\rm SU}_{1,1}(\/\mathbb C\/) &\equiv \hbox{\rm SU}(1,1)  
\qquad &\ \hbox{\rm SO}_{\rm o}(1,3)\ \leftrightarrow \ {\rm SL}(n,\mathbb C) \\
\mathbb H \quad&  \hbox{\rm SO}_{\rm o}(1,4) &\quad \leftrightarrow \quad& \hbox{\rm SU}_{1,1}(\/\mathbb H\/) &\equiv {\rm Sp}(1,1) 
\qquad &\ \hbox{\rm SO}_{\rm o}(1,5) \ \leftrightarrow \ {\rm SL}(n,\mathbb H)\\
\mathbb O \quad& \hbox{\rm SO}_{\rm o}(1,8) &\quad \leftrightarrow \quad& \hbox{\rm SU}_{1,1}(\mathbb O)  
&\qquad &\ \hbox{\rm SO}_{\rm o}(1,9) \ \leftrightarrow \ {\rm SL}(n,\mathbb O)
\end{array}
$$
The first three cases, real, complex and quaternionic  
($\mathbb R$, $\mathbb C$, $\mathbb H$), 
are considered in the following sections, while the octonions, $\mathbb O$,
will be analyzed elsewhere. 
This association is alternative to relation  
${\rm SL}(n,\mathbb F) \leftrightarrow {\rm SO}(1,\! n\!+\!1)$ shown above on the right side
\cite{Su}, often recalled in the context of supersymmetry \cite{KT, BH}.

~

\noindent \textbf{Example:}  To see the simplicity of the ``menhir calculus''   
consider two orthogonal velocities and the corresponding menhirs  
given as complex numbers: 
$$
v_1 =\frac{4}{5} , \quad  v_{2} =\frac{3i}{5}  \qquad  \   \Rightarrow \  \qquad    \varepsilon _1 =\frac{1}{2} ,\quad  \varepsilon _2 =\frac{i}{3} 
$$ 
Using \eqref{eq:intro11}, one finds the that the composition of the two implied boosts corresponds 
the menhir $\varepsilon=\varepsilon_1\moplus \varepsilon_2$ and velocity $v=v_1\oplus v_2$:
$$
              \varepsilon =\frac{1/2+i/3}{1+1/2\cdot i/3}  = \frac{20+9i}{37} 
\qquad\Rightarrow\qquad
 v = \frac{2\varepsilon }{1+\bar{\varepsilon }\varepsilon }   % = \frac{4(5+3i)}{25} 
                                                                                                =\frac{4}{5} +\frac{9}{25}i
$$ 
Thus velocity's direction is that of $5+3i$, and the speed is $4\sqrt{34}/25  \approx 14/15$.
%%%%\footnote{Use the famous ancient formula for rational approximation of square roots. } 
The rotational part of the composition of the boosts is 
$$
\rho = \frac{1+{\tfrac{1}{2}} {\tfrac{i}{3}} }{1-{\tfrac{1}{2}} {\tfrac{i}{3}} } = \frac{6+i}{6-i} = \frac{35+12i}{37} 
$$ 
which indicates that the angle of rotation to be $\theta = \arccos(35/37) \approx 19^\circ$.  
%(Remark: the fact that (8,15,17) form a Pythagorean triple is not accidental.)  
%================= FIGURE 1
\begin{figure}[h]
\centering
\includegraphics[scale=1.2]{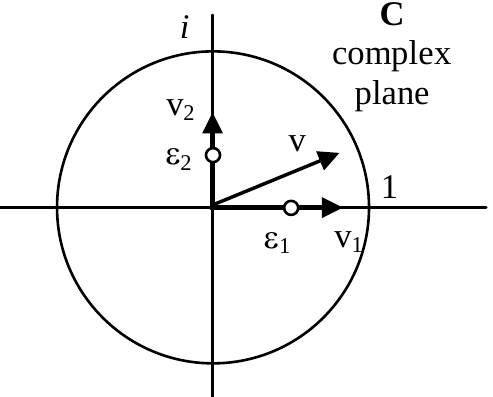}
\caption{\small Example.  The ``sum'' of velocities does not coincide with the sum of vectors.}
\label{fig:introexample}
\end{figure}

\begin{center}
* \qquad *\qquad   *
\end{center}

%For a general material on special relativity see \cite{Ri}. 
An alternative connection between the spinors and the Lorentz group was introduced in \cite{Eb} and popularized in \cite{PR}. 
For analysis of relativistic composition of velocities in terms of a grupoid see \cite{Oz1}.  
%Non-associativity of the velocities was addressed in \cite{MG, Un3}.  

\newpage

%------------ 4 ------------------------------------------------------------------------------------------------------------------------------------
\section{Reversions and unitary groups}

Here  we review the concept of reversion groups following \cite{jk-porism}. 
Let $D\subset E\cong \mathbb R^n$ be a unit disk and 
$K=\q D$ be the unit sphere in the Eucludean space $E$.

\begin{figure}[H]
\centering
\includegraphics[scale=1.1]{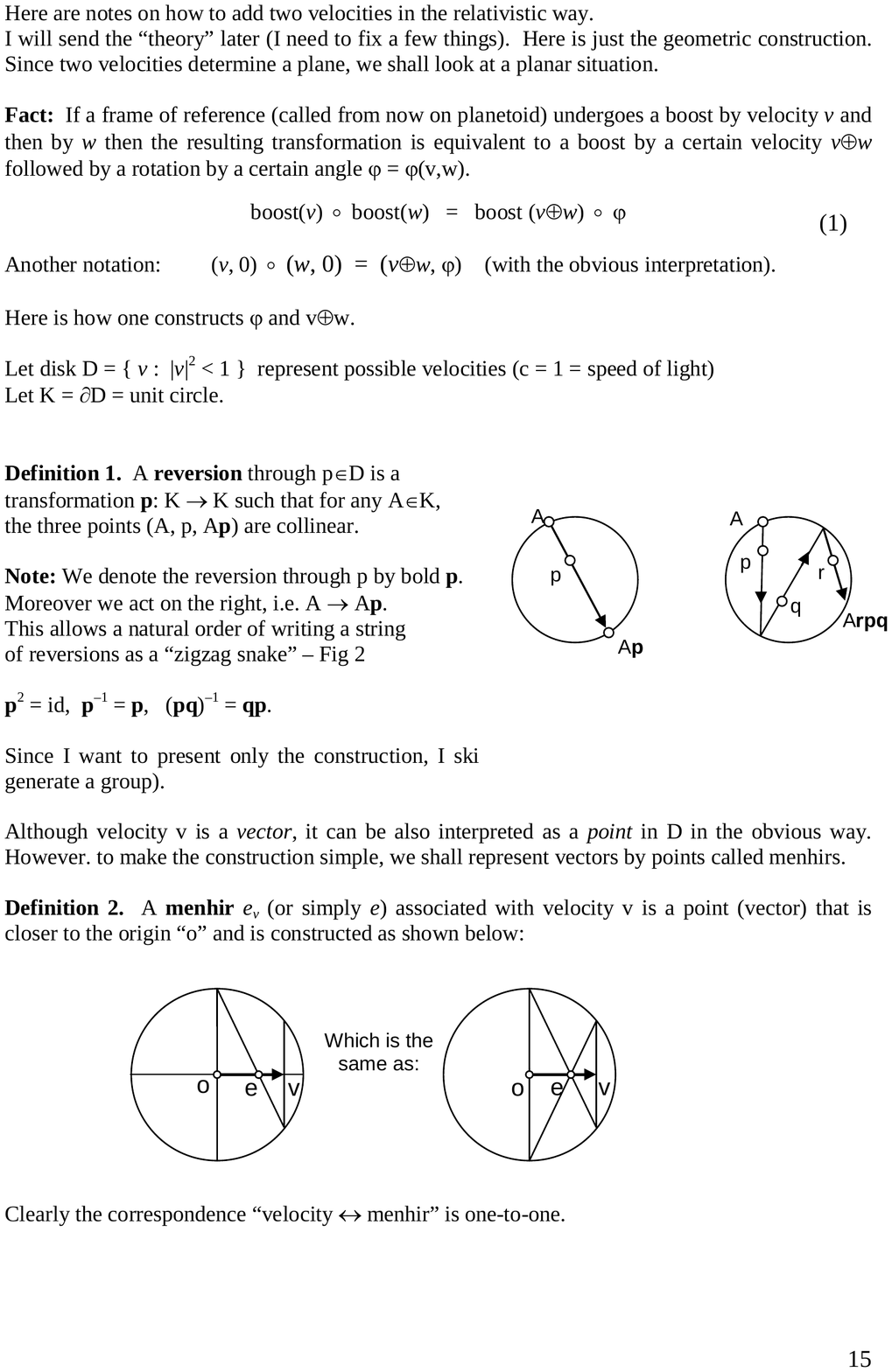}
\caption{\small Definition of reversion (left) and a composition of reversions (right)}
\label{fig:reversions2}
\end{figure}

\begin{definition}
A {\bf reversion} of a sphere $K$  through a point $p\in D$ is a map 
$$
\mathbf p: K\to K: A\mapsto A\mathbf p
$$ 
such that  points $(p,A,A\mathbf p)$ are collinear and $A\not= A\mathbf p$.
See Figure \ref{fig:reversions2} left  for illustration.
\end{definition}

%%================= FIGURE
%\begin{figure}[H]
%\centering
%\includegraphics[scale=.91]{C-reversion}
%\caption{\small Definition of reversion}
%\label{fig:reversion}
%\end{figure}

Reversions may be composed. In particular,  $\mathbf p^2 = \rmId$. 
They generate a group that we shall denote
$$
\hbox{Rev}(n) = \hbox{gen}\,\{\mathbf p\,|\, p\in D\}
$$
where $n=\hbox{dim}\, D$.
The subgroup of elements that consists of the composition of an even number of reversions will be denoted by 
$
\hbox{Rev}_{\rm o}(n) \subset \hbox{Rev}(n) 
$.

%================= FIGURE 1

In the special case of $n=2$, we may employ complex numbers and represent
$D$ and $K$ as unit disk and circle at the origin, $D= \{\varepsilon \in \mathbb C \,|\, |\varepsilon|<1\}$
and $K= \{z \in \mathbb C \,|\, |\varepsilon|=1\}$.

\begin{proposition}
A reversion through point represented by a complex number $\varepsilon\in\mathbb C$ can be then represented by a 
linear fractional transformation
%$$
\begin{equation}
\label{eq:bbi}
%\hbox{[reversion]}\qquad
       z \quad \rightarrow \quad z'  
   \ = \ \frac{z-\varepsilon }{\bar{\varepsilon }z-1} 
   \ = \  \left[\begin{array}{cc} {1} & {-\varepsilon} \\ {\bar{\varepsilon}}& {-1} \end{array}\right]\cdot z
\end{equation}
%$$
where the matrix on the right side is the standard notation for such maps.
\end{proposition}

One simply checks that this map preserves $K$ and that the image $z'$ is collinear with $\varepsilon$ and $z$
by verifying that $(z\varepsilon)(z'-\varepsilon)\in \mathbb R$.
\\

As an example of an even-order reversion, consider the following useful case: 
%Since reversion through $0$ is simply a change of sign of $z\in K$, we get

\begin{corollary}
\label{thm:tworev}
The composition of reversion through the origin followed by reversion through point $\varepsilon \in D$
has the following form
%$$
\begin{equation}
\label{eq:bibi}
              z \quad \rightarrow \quad   \frac{(-z)-\varepsilon }{\bar{\varepsilon }(-z)-1}  \ =\  \frac{z+\varepsilon }{\bar{\varepsilon }z+1} 
\ = \  \begin{bmatrix} 1&\varepsilon \\
\bar\varepsilon & 1\end{bmatrix} \cdot z
\, .
\end{equation}
%$$
\end{corollary}

~\\
\noindent
{\bf Group characterization of the matrices.}
The \eqref{eq:bbi} and \eqref{eq:bibi} suggest that the groups of reversions may be represented by matrix groups.
\begin{proposition}
The groups of reversions in dimension 2 are isomorphic to the special pseudo-unitary groups, namely:
\begin{equation}
\label{eq:}
\begin{aligned} 
                \Rev(2) &\cong  \hbox{\rm PSU}^{\pm} (1,1)  
\\
               \Rev_{\rm o}(2) &\cong  \hbox{\rm PSU}(1,1)  
\end{aligned}
\end{equation}
\end{proposition}

\noindent
{\bf Proof:}
The pseudo-unitary group $U(1,1)$ consists of matrices that preserve the inner product given by 
$$
g =  \left[\begin{array}{cc} 1 & 0 \\ 0 & -1 \end{array}\right]
$$
in the sense that $A^*gA=g$.  One can identify two special subgroups:
\begin{equation}
\label{eq:zuzu}
\begin{aligned} 
                  \hbox{\rm SU}^{\pm} (1,1)  & \ = \  \{\;   A\in U(1,1)  \;\big| \; \det A= \pm 1 \}
\\
                   \hbox{\rm SU}(1,1)  & \  =  \  \{\;   A\in U(1,1)  \;\big| \; \det A= 1 \}
\end{aligned}
\end{equation}
Clearly, $\hbox{\rm SU}(1,1) \subset \hbox{\rm SU}^{\pm} (1,1)$.
In the case of $\det A=1$ the matrices are of type:
$$
A= \left[\begin{array}{cc} a & b \\ \bar b & \bar a \end{array}\right] \,,
\qquad  |a|^2-|b|^2 =1 \,,
\qquad a,b\in \mathbb C
$$ 
Under projectivization this group becomes $\hbox{\rm PSU}(1,1)$.  
However it is very useful to use matrices that are scaled so that the connection with geometry (and later relativity) is simple.
Hence we start with the the group 
$\mathbb R_+\times \hbox{\rm SU}(1,1)$, which 
consists of matrices 
$$
A= \left[\begin{array}{cc} a & b \\ \bar b & \bar a \end{array}\right] \,,
\qquad  |a|^2-|b|^2 >0 \,,
\qquad a,b\in \mathbb C
$$
Note that the matrices of %\eqref{eq:bbi} and
\eqref{eq:bibi} are of this type.
Under projectivization,  the groups are identified, 
$P(\mathbb R_+\times \hbox{\rm SU}(1,1)) \cong \hbox{\rm PSU}(1,1)$.
The same goes for odd-order reversions for which the matrices are of form 
$$
\left[\begin{array}{cc} a & -b \\ \bar b & -\bar a \end{array}\right] 
\quad = \quad 
\left[\begin{array}{cc} a & b \\ \bar b & \bar a \end{array}\right] 
\left[\begin{array}{cc} 1 & 0 \\  0 &  -1 \end{array}\right] 
$$ 
with similar rules for $a$ and $b$. 
\QED
\\

%If we thing of the latter group as a fiber bundle $\mathbb R_+\times SU(1,1) \to PSU(1,1)$, then the matrices of equation (...)
%are just taken from a convenient section of the bundle.
We end with noting that every element of our even version of the group may be decomposed:
$$
\left[\begin{array}{cc} a & b \\ \bar b & \bar a \end{array}\right] = 
\left[\begin{array}{cc} a & 0 \\ 0 & \bar a \end{array}\right] \; 
\left[\begin{array}{cc} 1 & b/a \\ (b/a)^* & 1 \end{array}\right]  
$$
(Note that $a\not= 0$ since $|a|^2>0$).
\\

This establishes the isomorphism of the bottom right and central part of diagram in Figure \ref{fig:bigdiagram} 
for the 2-dimensional case and complex numbers.
A similar correspondence for $\dim D= 4$ and quaternions will be discussed in Section \ref{sec:algebra}.

\newpage 

%---------------------- 3    --------------------------------------------------------------------------------------------------------------------------
\section{\textit{As above, so below} -- Stonehenge applied}

{\bf Notation:}
\begin{enumerate}
\item
${\bf M}\equiv \mathbb R^{1,n}$ --- Minkowski vector space with an inner product $G$ {\small  of signature $(1,n)$.}
\\[-21pt]
\item
$\Lambda \equiv \hbox{\rm SO}_{\rm o}(1,n)$ --- {\bf Lorentz group}, the connected component of 
the symmetry group  $\hbox{\rm SO}(1,n)$ of $M$.
\\[-21pt]
\item
${\bf M}_F$ --- future unit hyperboloid consisting of the future-oriented unit vectors; 
a single sheet from the 2-component set $\{\mathbf w\in  \mathbb R^{1,n} \;|\; |\mathbf w|^2=1\}$.
\\[-21pt]
\item
$\mathbb R^{1,n}_0$  --- the {\bf light cone}, the subset of isotropic vectors of $\mathbb R^{1,n}$,  $G(w,w)=0$.
\\[-21pt]
\item
${\rm P}\mathbb R^{1,n}_0$  ---  the {\bf celestial sphere}, the projective space ${\rm P}\mathbb R^{1,n}$ restricted to the light cone.
Topologically,  ${\rm P}\mathbb R^{1,n}_0\equiv S^{n-1}$.
\\[-21pt]
\item
{\bf Tempus} --- a unit future-oriented time-like vector $\mathbf T \in M_F$. 
It determines an {\bf observer}, that is the split ${\mathbf M} \cong (\rmspan \mathbf T)\oplus \mathbf T^\bot$ 
where $\mathbf T^\bot = \{\mathbf w\in M \,|\, \mathbf w \bot \mathbf T \,\}$ \ is the associated perpendicular subspace (instantaneous space).
\\[-21pt]
\item
{\bf Lab} --- an $n$-dimensional Euclidean space $\mathbf E$ 
%with an inner product $G_\mathbf E$ 
together with a linear injective isometry $\lambda:\mathbf  E\to \mathbf M$.
%such that the induced map equates the inner products, $\lambda^*G = G_E$.
The embedding determines $\mathbf T_\lambda  \in \mathbf M_F$ such that $\mathbf T_\lambda\bot\lambda(E)$.
%The isometry group $\Lambda$ acts on the set of all labs $\{\lambda:E\to M\}$ by   $G^*(\lambda) = G\circ \lambda$
\end{enumerate}

\begin{figure}[H]
\centering
\includegraphics[scale=.79]{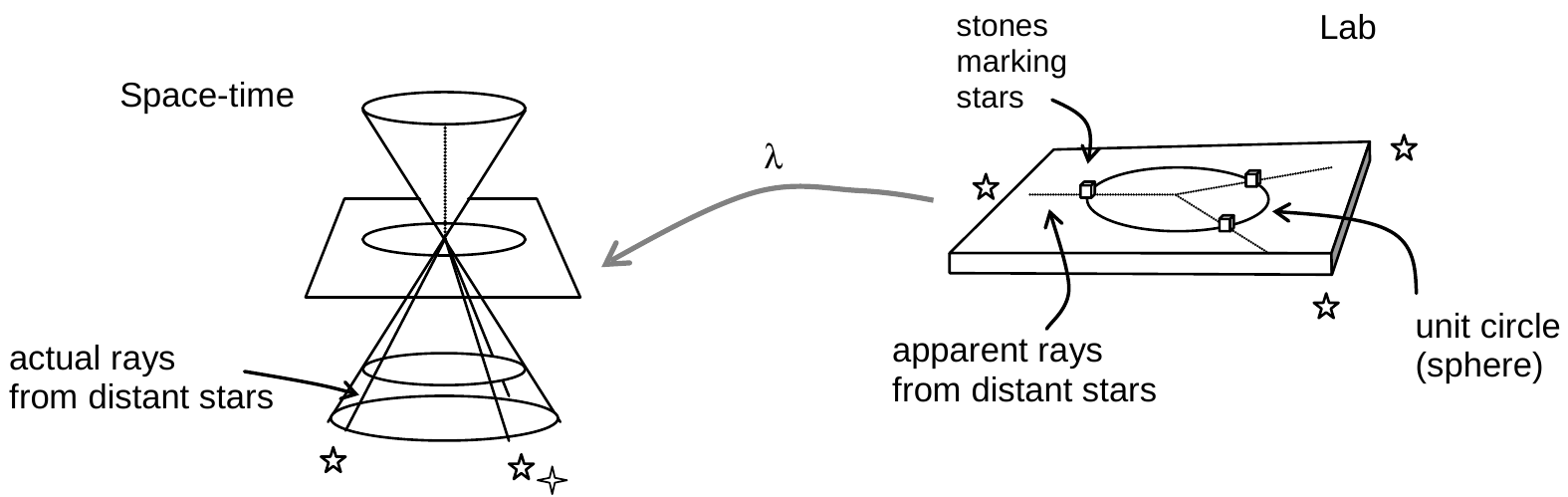}
\caption{\small Building a relativistic ``Stonehenge''.}
\label{fig:Stonehenge}
\end{figure}

\noindent 
\textbf{Making sense of ``adding velocities''.}
A lot of conceptual trouble may be avoided by introducing a notion of a ``lab'',
namely a reference Euclidean space $\mathbf E\cong \mathbb R^n$ together with injective isometry to the Minkowski vector space 
%$$
\begin{equation}
\lambda: \mathbf E\ \to\  \mathbf M
\end{equation}
%$$
One may think of a flat  ``planetoid'' like one in Figure \ref{fig:Stonehenge}
as its intuitive  metaphor. 
Suppose we can control the behavior of this planetoid:  
we may rotate it or give it a boost or in general,   
send the image of $\mathbf E$ to a new orientation in $\mathbf M$, 
so that $\lambda$ is replaced by a new embedding $\lambda^\prime=g\circ\lambda$,
which is a composition of the original $\lambda$ with an element of the group $g\in \Lambda$.
In particular, a boost by velocity $\mathbf v\in \mathbf E$ is understood as a map 
$B_{\lambda,\mathbf v}\in \Lambda$,
%defined by bivector $B_{\lambda,\mathbf v}$ proportional to $\lambda(\mathbf v)\wedge \mathbf T_{\lambda}$,
%sending the image of $E$ to a new orientation in $M$, that is to $\lambda$ is replaced by $B_{\mathbf v}\circ\lambda$.
%(the proportionality factor, easy to calculate, is not essential for this discussion).
for simplicity denoted 
$B_{\mathbf v}$. 
such that the new ``tempus" admits decomposition $\mathbf T_{\lambda^\prime} = \mathbf T_{\lambda}+\lambda(\mathbf v)$,
and no rotation appears, i.e., $(\mathbf T_{\lambda}\wedge \mathbf T_{\lambda^\prime})^\bot$ remains fixed under the boost.  
\\

We will make sense of adding velocities ``$\mathbf v\oplus \mathbf w$''  in the space $\mathbf E$
by the means of \textbf{planning}. 
Given an ordered pair of vectors, $\mathbf v, \mathbf w\in E$
%inside a unit disk.  
we execute two boosts: first $B_{\mathbf v}$ by velocity vector $\mathbf v$
and then $B_{\mathbf w} =B_{\mathbf w, B_{\mathbf v}\circ\lambda}$.
(Note that vector $\mathbf w$ is the \textit{same} vector in $\mathbf E$ 
but \textit{different} vector in space-time, namely $B_v(\lambda (\mathbf w))\in\mathbf M$.)

%================= FIGURE
%-----------------------
%
\noindent 
%How do we perceive a regular rotation? -- no mystery here.
 
One can reach the same result by a single boost by a velocity 
denoted $v \voplus w\in E$  followed by a certain rotation denoted $R_{v,w}$ :  
%$$
\begin{equation}
\label{eq:BBRB}
              B_w   \circ B_v  =  R_{v,w}\circ  B_{v\voplus w}
\end{equation}
%$$ 
%Thus the composition of velocities
The velocity $v\voplus w$ is viewed as the ``composition of velocities''  and it is only a part of the story.  
We are interested in finding the formulas for both components, 
the effective vector $v\voplus w$ and the angle and axis of rotation $R_{v,w}$.

%
%\noindent \textbf{Here is the idea}:
%This is represented by pullbacks of the action of the Lorentz group: $\lambda^{-1}g\lambda:E\to E$.

%================= FIGURE
\begin{figure}[h]
\centering
\includegraphics[scale=.6]{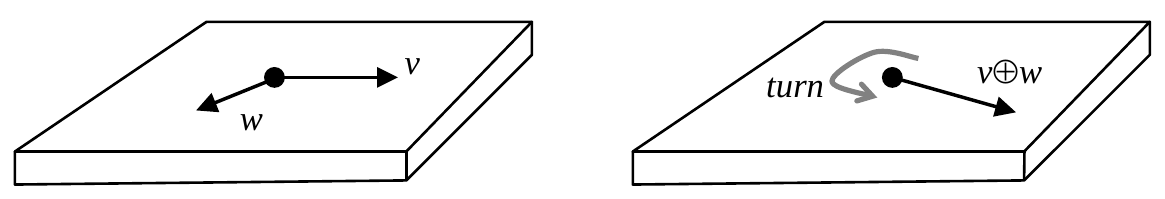}
\caption{\small Preplanned two boosts are marked on the ground of planetoid (lab).
They are equivalent to a single boost followed by rotation}
\label{fig:planetoid-boost}
\end{figure}

\noindent 
{\bf Cromlech.}
When we look at a starry sky at night -- we see a \textit{celestial sphere} made by rays from distant stars.  
Mathematically, we see points of the projective space restricted to the directions in the null cone, ${\rm P}\mathbb R^{1,n}_0$
(topologically equivalent to sphere $S^{n-1}$).  

%The action of the Lorenz group on $\lambda(E)$ will be represented as the action on a sphere in $E$.  
%in the observer's space We shall call it a ``Stonehenge setup''.   

To follow our 2-dimensional example of planetoid $\mathbb R^2$, 
imagine that a sky-watcher draws a unit circle $K\subset \mathbf E$, called \textbf{cromlech}.
Then he sets stones on this circle to mark particularly interesting stars 
(theoretically -- all points) as seen from the center on the horizon. 
%The circleand its points   \textbf{star stones}.  
He would perceive boosts and rotations as aberration of star positions, 
that is as conformal diffeomorphisms of the celestial sphere, 
or, equivalently, as diffeomorphisms of $K$.
Clearly, it generalizes to any dimension.
%They are stones representing the direction in which a hypothetical star is visible from the center of the cromlech.
%

%================= FIGURE
\begin{figure}[H]
\centering
\includegraphics[scale=.61]{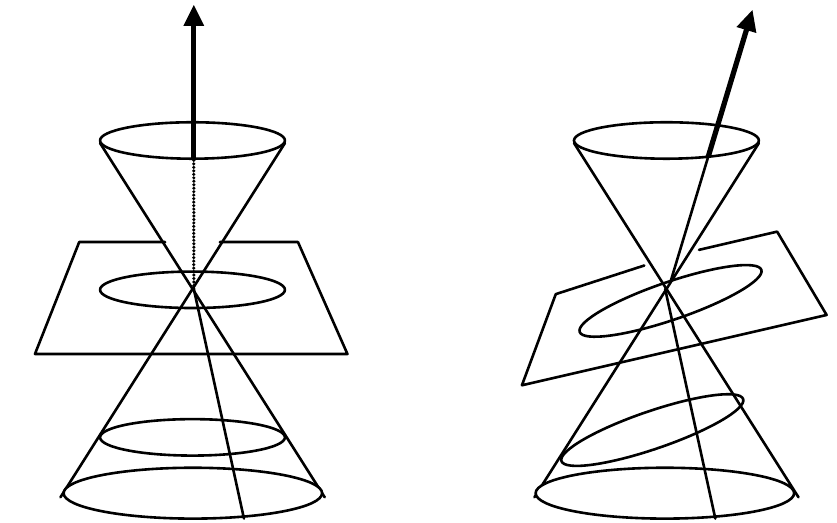}
\caption{\small ``Celestial sphere'' as a sphere in space}
\label{fig:celsphere}
\end{figure}
%----------

Here is a more precise description:  
For a particular observer (tempus) $\mathbf T = \mathbf T_\lambda$, 
we can map the celestial sphere to the unit sphere in an instantaneous space $\mathbf T^{\bot}$  
by cutting the light-cone by the hyperplane parallel to space $\mathbf T^{\bot}$  at the level  $(-\mathbf T)$  
and then projecting this intersection on the space $\mathbf T^{\bot}$ along $\mathbf T$
(see Figure \ref{fig:celsphere}).
%Negative ``$-\mathbf T$'' since  the rays of the distant stars are arriving from the past.
Mathematically, we get an $(n-1)$-sphere in $\lambda (\mathbf E)$:
$$
S^{n-1} \ \cong \ (\mathbf T^\bot - \mathbf T)\cap \mathbb R^{1,n}_0 + \mathbf T
$$
The map identifying $K$ with the celestial sphere is thus 
$$
\pi_\lambda:\ K\ \to \ \mathbb R^{1,n}_0 :\quad 
%\to \lambda{s} 
s \ \mapsto \ \hbox{span}\,\{\lambda(s) -\mathbf T_\lambda\,\}
$$

%$$
%\mathbb R^{1,n}_0 \ni a \quad \mapsto\quad \lambda^{-1}\circ \pi_T \in K\subset E
%$$

Let us now look how the boost-induced diffeomorphism of $K$ may be viewed geometrically.
Given velocity as a vector (point) $\mathbf v\in E$ inside the cromlech , $|v|^{2}< 1$,  
the associated boost  $B_{\mathbf v}$ 
will make the stars on the sky undergo an {\bf aberration}, they will shift towards the boost direction, 
except the star in front and behind that remain unaffected. 
%This aberration of stars may be presented as a diffeomorphism of the cromlech.

%\newpage

%================= FIGURE
\begin{figure}[h]
\centering
\includegraphics[scale=.81]{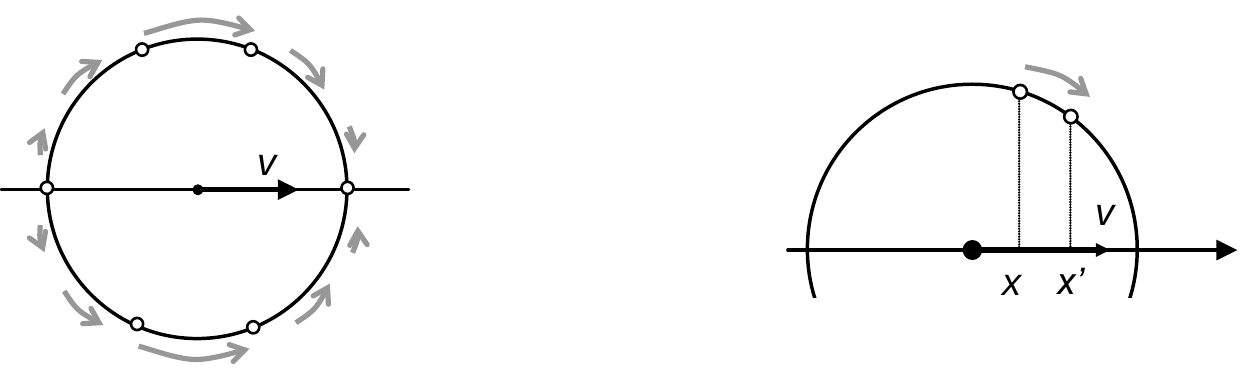}
\caption{\small {\bf Left:} 
A single boost boost makes a shift of stars in the direction of the velocity.\\
{\bf Right:} Shift of stars in cromlech due to a boost measured along the velocity axis.}
%\caption{\small {\bf Left:} Sci-Fi puzzle for scouts:  When you woke up, you noticed that the sky is deformed.  
%You learn that your planet has undergone a single boost in unspecified direction. Find the direction and the value of the velocity change.\\
%{\bf Right:} Shift of stars in cromlech due to a boost by velocity $v$ }
\label{fig:starpuzzle}
\end{figure}

\newpage
\noindent 
Here  is our initial simple observation:

\begin{proposition}
\label{thm:xv}
The stars represented in our the cromlech will change the positions so that 
their projections $x$ on the axis along the velocity is transformed to a new value $x'$ according to 
%$$
\begin{equation}
\label{eq:xv}
              x  \ \rightarrow \     x'  =  \frac{x+v}{1+vx} 
\end{equation}
%$$
\end{proposition}
\noindent \textbf{Proof:}   Denote  $s = \sinh \omega$,   $c = \cosh \omega$,  $t = \tanh \omega  = v$.  
The transformation of the stars on the celestial sphere corresponds to action of $A^{-1}$ on the vectors $[t = -1, x]$, where   
$$
         A  =  \left[\begin{array}{cc} {c} & {s} \\ {s} & {c} \end{array}\right] \,,
\qquad          
        A^{-1}  =  \left[\begin{array}{cc} {c} & {-s} \\ {-s} & {c} \end{array}\right]\,.
$$
\noindent Calculations may be done for the two-dimensional subspace span by $x$ and $t$ :
$$
           \left[\begin{array}{cc} {c} & {-s} \\ {-s} & {c} \end{array}\right]\left[\begin{array}{c} {-1} \\ {x} \end{array}\right]
\  = \   \left[\begin{array}{c} -c\!-\! sx  \\  s+cx \end{array}\right]
\ \dot{=}\  
           \left[\begin{array}{c} {-1} \\ {{\tfrac{s+cx}{s+sx}} } \end{array}\right]
 \quad \Rightarrow \quad    x'   =  \frac{cx+s}{sx+c} 
$$ 
Now, divide the numerator and denominator by $\cosh \theta$ and use the fact that $s/c = \tanh \theta  = v$    
to get Equation \eqref{eq:xv}.

%================= FIGURE
\begin{figure}[H]
\centering
\includegraphics[scale=.91]{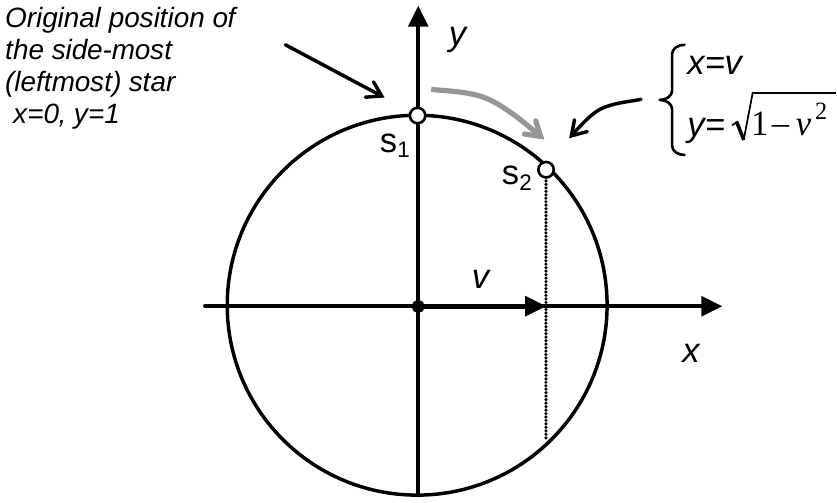}
\caption{\small Velocity of boost and the shift of the side-most stars in 2-dimensional case}
\label{fig:starv}
\end{figure}

%-------------------------------
\noindent \textbf{Corollary 3.2:} 
The two antipodal star stones that did not change their positions determine the axis of velocity $v$.  
The side-most stars will move to position such that the corresponding stones will be aligned with the velocity stone.  
See Figure \ref{fig:starv}.
\\
\\
{\bf Proof:} Substitute  $x=0$ for the side-most stars 
to \eqref{eq:xv},
and $x=\pm 1$ for the axis points.  \QED 

%\newpage

%\newpage

%------------ 4 ------------------------------------------------------------------------------------------------------------------------------------
\section{Celestial sphere, menhirs and star motion}

Now we show how to construct the deformation of the celestial sphere in a purely geometric method.
Let us go back to the ``planetoid'' setup: consider a lab 
$\mathbf E\cong \mathbb R^n$ embedded in the Minkowski space $\mathbf M\cong \mathbb R^{1,n}$.  
Denote $D\subset \mathbf E$ the disk of vectors of the norm not exceeding 1.  
They will represent velocities, $|\mathbf v|<1$, candidates for boosts.
The unit sphere $K=\q D$ will be called ``cromlech'' for the reasons explained. Its center is denoted $O$.

\begin{figure}[h]
\centering
\includegraphics[scale=.81]{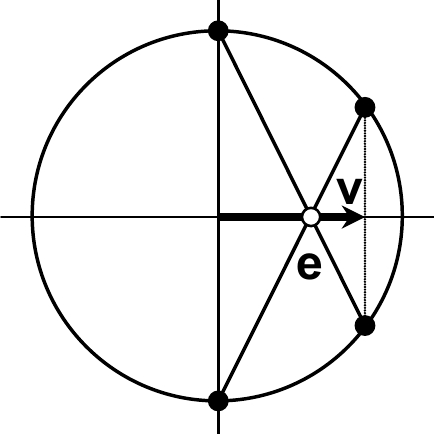}
\caption{\small Position of menhir $\mathbf e$ for a given velocity}
\label{fig:menhir}
\end{figure}

\begin{definition}
A {\bf menhir} associated to velocity $\mathbf v$ is a vector $\mathbf e\in D$  
the construction of which is shown in Figure \ref{fig:menhir}.
The corresponding one-to-one map 
$$
\mu: D\ni\mathbf v \ \longrightarrow \ \mathbf e \in D 
$$
will be called menhir map.  Vectors $\mathbf e$ will geometrically interpreted as a point in $D$.
\end{definition}

%================= FIGURE

\begin{proposition}
The relation between  velocity $\mathbf v$ and the associated menhir $\mathbf e=\mu(\mathbf v)$ is
%$$
\begin{equation}
\label{eq:ve}
%           \mathbf e \quad \ontop{\mu} \longrightarrow \quad 
 \mathbf  v  = \frac{2\mathbf e}{1+e^{2}  }
\end{equation}
%$$
where $e=|\mathbf e|$.
\end{proposition}

\noindent
{\bf Proof:}
Use similarity of triangles, Figure \ref{fig:menhir}, to get $e:1 = v:(1+\sqrt{1-v^2})$.  Solve for $v$. \QED
\\

\noindent 
{\bf Remark:} The relation between the $v$ and $e$ (absolute values) can be presented in the following elegant form
%$$
\begin{equation}
\frac{1-v}{1+v} \ = \ \left(\frac{1-e}{1+e}\right)^2
\label{eq;nice}
\end{equation}
%$$ 
When solved for $\mathbf v$, it resolves in \eqref{eq:ve}, when resolved for $\mathbf e$, one gets
the inverse relation
%$$
\begin{equation}
                \mathbf e=\frac{\mathbf v}{1+\sqrt{1-v^{2} } }\,.
\end{equation}
%$$ 
Just for completeness, here are some other forms of these relations
%$$
\begin{equation}
v=\frac{(1+e)^2-(1-e)^2}{(1+e)^2+(1-e)^2}
\ \qquad\ 
e=\frac{\sqrt{1+v}-\sqrt{1-v}}{\sqrt{1+v}+\sqrt{1-v}}
\end{equation}
%$$

Now we can state our first important result.

%---------------------
\begin{theorem} 
\label{thm:boost}
A boost by velocity $\mathbf v$ causes deformation of the celestial sphere $K$ in the way that coincides with 
a composition of two reversions: 
through the origin followed and the menhir $\mathbf e=\mu(\mathbf v)$.
That is, a star at position $A\in K$ will shift to position 
$$
       A \ \mapsto \ A' = A\mathbf o \mathbf e
$$
 \end{theorem}

\begin{figure}[H]
\centering
\includegraphics[scale=.81]{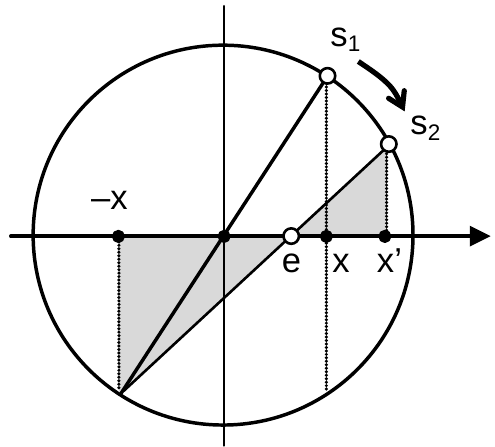}
\caption{\small Similar triangles -- proof of Theorem \ref{thm:boost}}
\label{fig:menhirproof}
\end{figure}

\noindent
\textbf{Proof:} Pick a point $s_1\in K$ and construct its image $s_2$ via the stated composition of reversions $\mathbf {eo}$, 
Figure \ref{fig:menhirproof}.
Project the points on the axis defined by the velocity $\mathbf v$, interpreted as a number axis.
The two shaded triangles are similar and by Thales' theorem, 
$$
\sqrt{1-x'^{2} } :  (x'-e) = \sqrt{1-(-x)^{2} }: (e+x)\,.
$$  
After squaring both sides and organizing as a polynomial equation, notice that one may factor out a term $(x+x')$.
The remaining terms are linear in $x'$ and we readily get 
$$
         x'=\frac{x+\frac{2e}{1+e^{2} } }{1+x\frac{2e}{1+e^{2} } } \,.
$$ 
This, comparing with \eqref{eq:ve} and \eqref{eq:xv}, gives the result. 
\QED

%================= FIGURE

\begin{figure}[H]
\centering
\includegraphics[scale=.81]{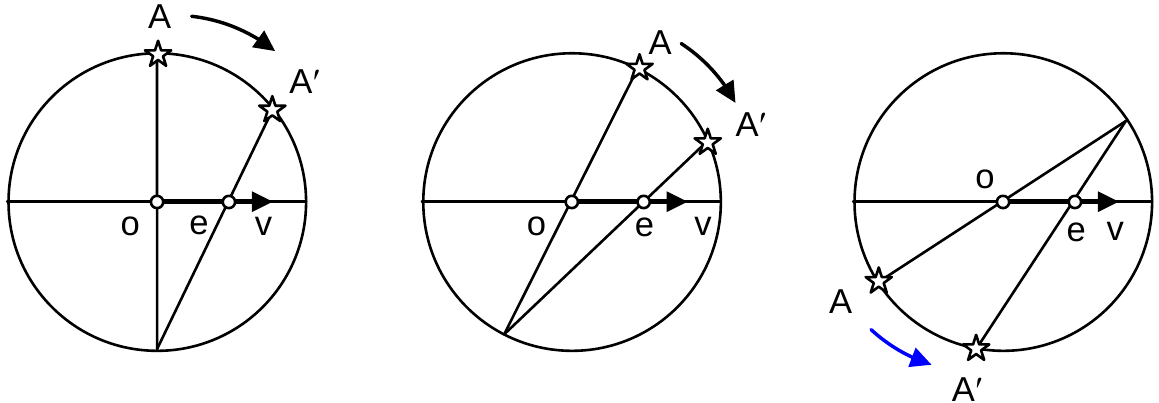}
\caption{\small Shift of stars constructed with a menhir}
\label{fig:starshift}
\end{figure}

\noindent 
We have a simple geometric method of describing the action of the Lorentz group on the celestial sphere.
represented by the cromlech --- see Figure \ref{fig:starshift}. 
\\

%================= FIGURE

\noindent
{\bf Towards geometric method of composition of velocities}.
Now we see how  two boosts may be represented via deformation of the celestial sphere $K\to K$:
Suppose the first performed boost is along velocity $v$, followed by the boost along $w$.
Let the corresponding menhirs are $e$ and $f$,  $\mu(e) = v$ and $\mu(f) = w$.
Then the star shift is 
$$
A \ \mapsto \ A' = A\mathbf o\mathbf e\mathbf o \mathbf f
$$
%as shown in Figure x.
This will be  simplified to a form that indicates the ``sum of velocities'' $v\oplus w$ more directly 
in Section \ref{sec:mehirvelocities}.  
\\\\

\begin{figure}[h]
\centering
\includegraphics[scale=.97]{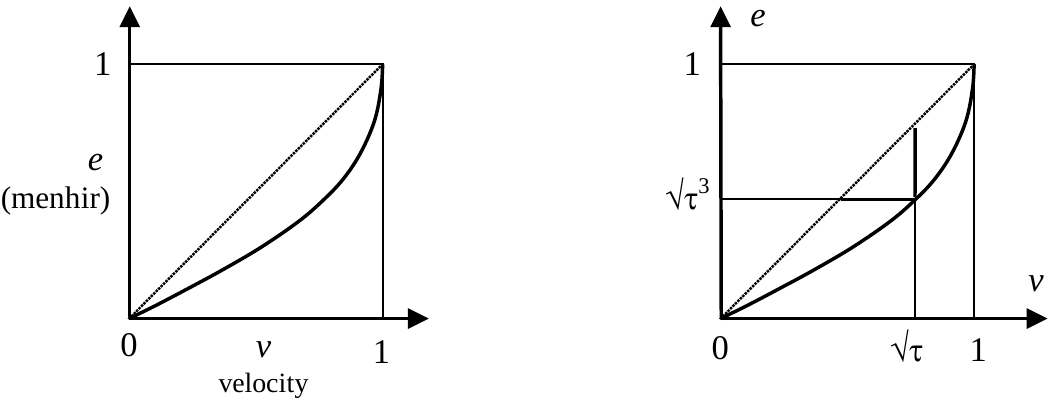}
\caption{\small Position of the menhir as a function of velocity}
\label{fig:function}
\end{figure}

%------------------------------------------------------------
\noindent \textbf{Remark (Phi in the sky):}   
The relation between the velocity and the position of the menhir $\mathbf e$ a non-uniform monotone map, 
shown in Figure \ref{fig:function} (for absolute values).   
The end-points coincide: the null velocity $v= 0$ corresponds to $e = 0$ 
and $v  = 1$ to $\varepsilon = v=1$.
The rule of thumb is that for the small values we have $\varepsilon \approx v/2$. 
%The menhir stone moves fast as the velocity reaches the speed of light 1.

%================= FIGURE

\noindent
One may ask: For what velocity the discrepancy between $v$ and $e$ is greatest. 
A quick excursion in calculus reveals that it happens for 
$$
v = \varphi^{-1/2}   \qquad \hbox{where}\qquad \varphi = (1+\sqrt{5})/2
$$
is the golden ratio! 
%So it is ``phi in the sky'' after all, not -- ``pi''\dots   
This implies that in the situation of the greatest discrepancy the menhir cuts the segment representing 
the velocity in the golden proportion.
Indeed, denoting the ``golden cut'' by $\tau = 1/\varphi = (\sqrt{5} - 1)/2$, 
we get velocity $v = \sqrt{\tau}$ % \approx  .79$ 
and the menhirs position $e = \sqrt{\tau^3}$,   %\approx .49$.
and therefore $v:e=\varphi$. 
%The difference is also related to the golden ratio: $\sqrt{\tau}-\sqrt{\tau}^3 = \sqrt{\tau}^5$. 

~

%---------------------------------------------------------
\begin{equation}
\label{eq:smalldiagram}
\begin{tikzpicture}[baseline=-0.8ex]
    \matrix (m) [ matrix of math nodes,
                         row sep=3em,
                         column sep=3em,
                         text height=4ex, text depth=3ex] 
 {
           \quad \dgaction{\Rev_{\rm o}(n)}{S^{n-1}}   \quad   
          &\quad  \dgaction{\hbox{\rm SO}_{\rm o}(1,n)}{\mathbb R^{1,n}_{0}} \quad    \\
           \quad \dgaction{\Rev(n)}{S^{n-1}}   \quad   
          &\quad  \dgaction{\hbox{\rm SO}_{\rm o}^{+}(1,n)}{\mathbb R^{1,n}_{0}} \quad     \\
  };
    \path[stealth-stealth]   (m-1-1) edge node[above] {$1:1$}  (m-1-2);
%    \path[stealth-stealth]        (m-1-2) edge node[above] {$1:1$}  (m-1-3);
%    \path[stealth-stealth]        (m-1-3) edge node[above] {$1:1$}  (m-1-4);
    \path[stealth-stealth]   (m-2-1) edge node[above] {$1:1$}  (m-2-2);
%    \path[stealth-stealth]        (m-2-2) edge node[above] {$1:1$}  (m-2-3);

    \path[stealth-]        (m-2-1) edge node[right] {$\subset$}  (m-1-1);
    \path[stealth-]        (m-2-2) edge node[right] {$\subset$}  (m-1-2);

\node at (m-1-1.north) [above=-1pt, color=black] {\smalll\sf {reversions}};
\node at (m-1-2.north) [above=-1pt, color=black] {\smalll\sf {relativity}};

\node at (m-1-1.north) [above=7pt, color=black] {\smalll\sf {GEOMETRY}};
\node at (m-1-2.north) [above=7pt, color=black] {\smalll\sf {PHYSICS}};

\end{tikzpicture}   
\end{equation}

\newpage
%---------------------- 5 --------------------------------------------------------------------------------------------------------------------------
\section{Algebraic version of the ``menhir calculus''}% and the composition formula }
\label{sec:algebra}
In this section we derive algebraic representation of the geometry developed in the previous section.  
We will consider in succession three division algebras $\mathbb F = \mathbb R, \mathbb C, \mathbb H$, 
which describe 1, 2, 4- dimensional space (the standard physical 3D space is contained in $\mathbb H$).  
Menhirs, as points inside the disk represent indirectly velocities via the map (from menhirs to velocities):
$$
               \mu: \;  D^{n} \rightarrow D^{n}:  \  v \rightarrow  \varepsilon
$$
defined by relation
%$$
\begin{equation}
\label{eq:eev}
                          v  = \frac{2\varepsilon}{1+|\varepsilon|^{2} } \qquad v,\varepsilon\in\mathbb F
\end{equation}
%$$

\paragraph{A. One-dimensional case and $\mathbb R$}

~\\\\
\noindent Recall that collinear velocities add up according to the Poincar\'e relativistic formula:
%$$
\begin{equation}
\label{eq:vplusw}
                  v\voplus w\ =\  \frac{v+w}{1+vw}
\end{equation}
%$$
This scalar algebraic formula turns the segment $D= (-1,1)$ into an Abelian group (the inverse of $v$ is $-v$, the neutral element is $0$).  
We may translate this to the algebraic expression for behavior of the menhir values of $e$.  Result is somewhat unexpected:

%----------------------------------- 
\begin{theorem}
\label{thm:epluse}
The menhirs corresponding to collinear velocities obey the same addition formula as the vectors of velocity, 
i.e., they ``add up'' {\it \`a la} Poincar\'e: 
%$$
\begin{equation}
\label{eq:epluse}
              e_{1} \moplus e_{2} \ =\  \frac{e_{1} +e_{2} }{1+e_{1} e_{2} }
\end{equation}
%$$
That is, we have a group isomorphism $\mu: (D,\voplus)\to(D,\moplus)$:
$$
\mu(\mathbf v) \moplus \mu(\mathbf w) \ = \ \mu(\mathbf v\voplus \mathbf w)
$$
%In other words, the map $\mu$ is an automorphism of the Poincar\'e addition group $(I, \voplus)$.
\end{theorem}
~\\
\textbf{Proof:}  Substitute \eqref{eq:eev} to \eqref{eq:vplusw} and simplify
$$
\begin{aligned}
v_{1} \voplus v_{2}   &=    \frac{v_{1} +v_{2} }{1+v_{1} v_{2} }    
                                =    \frac{\frac{2e_{1} }{1+e_{1}^{2} } +\frac{2e_{2} }{1+e_{2}^{2} } }{1+\frac{2e_{1} 2e_{2} }{(1+e_{1}^{2} )(1+e_{2}^{2} )} } 
\\[7pt]
%                              &=    \frac{2e_{1} (1+e_{2}^{2} )+2e_{2} (1+e_{1}^{2} )}{(1+e_{1}^{2} )(1+e_{2}^{2} )+4e_{1} e_{2} }    
%                               &\hbox{(cross-multiply) }
%\\[7pt]
                             &=  \frac{2(e_{1} +e_{2} )(1+e_{1} e_{2} )}{(e_{1} +e_{2} )^{2} +(1+e_{1} e_{2} )^{2} }        
                             & \hbox{(rearranging terms)}
\\[7pt]
                              &=  \frac{2\frac{e_{1} +e_{2} }{1+e_{1} e_{2} } }{\left(\frac{e_{1} +e_{2} }{1+e_{1} e_{2} } \right)^{2} +1}     
                                   &(\hbox{factor} \  (1+e_{1}e_{2})^{2}) 
\end{aligned}
$$
\noindent 

\noindent Comparing this with \eqref{eq:vplusw} gives the result.  \QED
\\
 
\noindent \textbf{Remark:}  Note that the map $\mu$ is the square in the sense of the group action,  $\mu(e) = e\moplus e$.  
In other words, the menhirs are the ``group square roots'' of velocities  (or rather ``halves'' -- if you prefer the additive terminology). 
%This result is the source of the geometrization of Poincar\'e addition given in the Introduction (see Figure \ref{fig:vdiagram}).  
%Since it holds for menhirs with their inversive geometry, it may be transferred to velocities because of the diagram \eqref{aaa}.
Note that due to the commutativity and associativity in this 1-dimensional case we have
$$
(e\boxplus e) \boxplus (f\boxplus f) = (e\boxplus f) \boxplus (e\boxplus f)
$$

%------------------------------------------------------------------------------
\paragraph{B. Two-dimensions and complex numbers $\mathbb C$}

~\\\\ 
Two-dimensional case is sufficient to illustrate composition of a pair of non-collinear velocities.  
We equip the space $\mathbf E$ 
(the ground of our 2-dimensional planetoid) with a complex structure, $\mathbb C\cong \mathbb R^2$.  
The choice of the real axis is inessential.  The cromlech is now $K= \{z\in \mathbb C \, : \, |z|^2=1\}$.

%---------------------
\begin{theorem}
\label{thm:ematrix2}
Under a boost, the points on the unit circle (stones of cromlech representing stars) are transformed
via complex linear fractional maps  
%$$
\begin{equation}
\label{eq:ematrix}
             z\quad  \rightarrow \quad      z' \ = \  \frac{z+\varepsilon }{\bar{\varepsilon }z+1}
                                            \ \equiv \ \left[\begin{array}{cc} {1} & {\varepsilon } \\ {\bar{\varepsilon }} & {1} \end{array}\right]\cdot z  
\end{equation}
%$$
where $\varepsilon$ is the complex number representing the menhir (reduced velocity) and $z\in K$ is 
a unit complex number. In particular, the map preserves $K$.
%(We use the standard dot notation for M\"obius action represented by a matrix.)
\end{theorem}

\noindent \textbf{Proof:}  
This follows directly fron Corollary \ref{thm:tworev}. 
%Following \cite{jk-porism},  
%a reversion through point represented by a complex number $\varepsilon\in\mathbb C$ can be described by a 
%fractional transformation
%$$
%\hbox{[reversion]}\qquad
%       z \ \rightarrow \  \left[\begin{array}{cc} {1} & {-\varepsilon} \\ {\bar{\varepsilon}}& {-1} \end{array}\right]\cdot z
%    =  \frac{z-\varepsilon }{\bar{\varepsilon }z-1} 
%$$
%One simply checks that this map preserves $K$ and that the image $z'$ is collinear with $\varepsilon$ and $z$
%by simply checking that $(z\varepsilon)(z'-\varepsilon)\in \mathbb R$.
%Now, since the reversion through the origin is simply a change of sign, we get
%$$
%              z \ \rightarrow \   \frac{(-z)-\varepsilon }{\bar{\varepsilon }(-z)-1}  = \frac{(z)+\varepsilon }{\bar{\varepsilon }z+1} 
%$$
%hence the claim.
 \QED  
\\

%\noindent \textbf{Remark 1:} We could rescale the matrix \eqref{eq:ematrix} to be in $SL(2,\mathbb C)$, 
%as one often does with the M\"obius transformations, but then we would lose its simple form.  
%\\

\noindent \textbf{Remark 1:}    Rotation is realized by $z\rightarrow  e^{i\varphi} z$.  
Thus boost given by $\varepsilon$ and followed by a rotation is of form
$$
      z \ \rightarrow \ z''  =  e^{i\varphi}\cdot \left[\begin{array}{cc} {1} & {\varepsilon } \\ {\bar{\varepsilon }} & {1} \end{array}\right] \cdot z  
                              %      =  \frac{e^{i\varphi } z+e^{i\varphi } \varepsilon }{\bar{\varepsilon }z+1} 
                                   =  e^{i\varphi } \frac{z+\varepsilon }{\bar{\varepsilon }z+1} 
$$

~

\noindent 
\textbf{Remark 2:} The product of a scalar and a matrix behaves in the following way: 
%$$
\begin{equation}
\label{eq:remarkmatrix}
         a\cdot \left[\begin{array}{cc} {\alpha } & {\beta } \\ {\gamma } & {\delta } \end{array}
       \right]\quad \equiv \quad 
        \left[\begin{array}{cc} {a\alpha } & {a\beta } \\ {\gamma } & {\delta } \end{array}\right]
         \quad \equiv \quad 
         \left[\begin{array}{cc} {\alpha } & {\beta } \\ {\gamma /a} & {\delta /a} \end{array}\right]\,. 
\end{equation}
%$$
(to reflect the meaning of the matrices as M\"obius action).
\\

\noindent Now, we are ready to see how a composition of boosts is represented in stone (menhir) realization in Argand plane.  Here is the result:

%----------------------------
\begin{theorem}
\label{thm:2Dee} 
{\bf [Master equation -- complex version].}  
Let $\varepsilon_1$, $\varepsilon_2 \in\mathbb C$ be two complex numbers that represent menhirs for two vectors $v_1, v_2\in\mathbb C$.   
The composition $B(v_2)B(v_1)$ of two pre-designed  boosts is equivalent to a single boost $B(v_1\voplus v_2)$ followed by a rotation, 
where the menhir representation of $ v\oplus w$ is 
%$$
\begin{equation}
\label{eq:2Dee1}
            \varepsilon_1 \moplus \varepsilon_2   =  \frac{\varepsilon _{1} +\varepsilon _{2} }{1+ \bar{\varepsilon }_1\varepsilon _2  }
\end{equation}
%$$
and the associated Thomas rotation $\varphi$ is twice the Arg$(1+\bar{\varepsilon }_{1} \varepsilon _{2} )$, 
or, in the form $\rho = e^{i\varphi}$, it is 
%$$
\begin{equation}
\label{eq:2Dee2}
       \rho = e^{i\varphi}  =   \frac{1+\bar{\varepsilon }_{1} \varepsilon _{2} }{1+\varepsilon _{1} \bar{\varepsilon }_{2} } 
\end{equation}
%$$
This can be gathered in this \textbf{C}-version of  the ``master equation'':
%$$
\begin{equation}
\label{eq:2Dee3}
\left[\begin{array}{cc} {1} & {\varepsilon _2 } \\ {\bar{\varepsilon }_2 } & {1} \end{array}\right]
\left[\begin{array}{cc} {1} & {\varepsilon _1 } \\ {\bar{\varepsilon }_1 } & {1} \end{array}\right]    
=     \frac{1+\bar{\varepsilon }_{1} \varepsilon _{2} }{1+\varepsilon _{1} \bar{\varepsilon }_{2} } \cdot 
         \left[\begin{array}{cc} {1} & {\frac{\varepsilon _{1} +\varepsilon _{2} }{1+\bar{\varepsilon }_{1} \varepsilon _{2} } } \\ 
                                          {\frac{\bar{\varepsilon }_{1} +\bar{\varepsilon }_{2} }{1+\varepsilon _{1} \bar{\varepsilon }_{2} } } & {1} \end{array}\right]
\end{equation}
%$$

\end{theorem}
\noindent
\textbf{Proof:}  Using the menhir representation \eqref{eq:ematrix}, a composition of two boosts gives the following chain of equations -- 
in which the matrices represent the M\"obius action on the cromlech
$$
\left[\begin{array}{cc} {1} & {\varepsilon _2 } \\ {\bar{\varepsilon }_2 } & {1} \end{array}\right]
\left[\begin{array}{cc} {1} & {\varepsilon _1 } \\ {\bar{\varepsilon }_1 } & {1} \end{array}\right] \cdot  z   
= 
\left[\begin{array}{cc} {1+\bar{\varepsilon }_{1} \varepsilon _{2} } & {\varepsilon _{1} +\varepsilon _{2} } \\ 
                                  {\bar{\varepsilon }_{1} +\bar{\varepsilon }_{2} } & {1+\varepsilon _{1} \bar{\varepsilon }_{2} } \end{array}\right] \cdot z   
=   
\frac{1+\bar{\varepsilon }_{1} \varepsilon _{2} }{1+\varepsilon _{1} \bar{\varepsilon }_{2} } \cdot 
                \left[\begin{array}{cc} {1} & {\frac{\varepsilon _{1} +\varepsilon _{2} }{1+\bar{\varepsilon }_{1} \varepsilon _{2} } } \\ 
                                            {\frac{\bar{\varepsilon }_{1} +\bar{\varepsilon }_{2} }{1+\varepsilon _{1} \bar{\varepsilon }_{2} } } & {1} \end{array}\right] \cdot    z
$$
where we took advantage of \eqref{eq:remarkmatrix}:   
the numerator is extracted from the upper row of the matrix, and the denominator from the lower row.  
\QED
\\

\noindent 
{\bf Remark:} 
Clearly \eqref{eq:2Dee1} is in general a non-commutative product.  
However, if $\varepsilon_1$ and $\varepsilon_2$ are collinear, then they are both in the form  $\varepsilon_k = \exp(i\varphi) e_k$  for some real $e_k$, 
$k=1,2$, which under substitution reduces  equations \eqref{eq:2Dee1}  to \eqref{eq:epluse}. 
\\
\\
{\bf Remark:}
The pairs $(D,\boxplus)$ and  $(D,\oplus )$ are not groups but only loops. 
They have a neutral element, $0$, and well-defined inverse, but are neither commutative nor associative. 
Yet we have the isomorphism 
$$
\mu(a) \oplus \mu(b) \ = \ \mu(a\boxplus b)
$$

The results are summarized by the following commutative diagram or relations between the reversions, matrices, 
and celestial action of the Lorentz group:

%\begin{corollary} The results can be summarized by the following homomorphisms and relations between the reversions, matrices, 
%and celestial action of the Lorenz group:
%
%
%
%---------------------------------------------------------
\begin{equation}
\label{eq:bigdiagramC}
\begin{tikzpicture}[baseline=-0.8ex]
    \matrix (m) [ matrix of math nodes,
                         row sep=3em,
                         column sep=2.8em,
                         text height=4ex, text depth=3ex] 
 {
           \quad \dgaction{\mathbb R_+\!\times \hbox{\rm SU}(1,1)}{S^1\subset\mathbb C}   \quad
          &\quad \dgaction{\hbox{\rm PSU}(1,1)}{S^1\subset\mathbb C}   \quad
          & \quad \dgaction{\Rev_{\rm o}(2)}{S^1\subset\mathbb R^2 \ }   \quad   
          &\quad  \dgaction{\hbox{\rm SO}_{\rm o}(1,2)}{\mathbb R^{1,2}_{0}} \quad    \\
             \quad \dgaction{\mathbb R_+\!\times \hbox{\rm SU}^{\pm}\!(1,1)}{S^1\subset\mathbb C}   \quad 
          & \quad \dgaction{\hbox{\rm PSU}^{\pm}\!(1,1)}{S^1\subset\mathbb C}   \quad 
          & \quad \dgaction{\Rev(2)}{S^1\subset\mathbb R^2 \ }   \quad   
          &\quad  \dgaction{{\rm SO}^\pm_{\rm o}(1,2)}{\mathbb R^{1,2}_{0}} \quad     \\
  };
    \path[-stealth]   (m-1-1) edge node[above] {$\pi$}  (m-1-2);
    \path[stealth-stealth]        (m-1-2) edge node[above] {$1:1$}  (m-1-3);
    \path[stealth-stealth]        (m-1-3) edge node[above] {$1:1$}  (m-1-4);
    \path[-stealth]   (m-2-1) edge node[above] {$\pi$}  (m-2-2);
    \path[stealth-stealth]        (m-2-2) edge node[above] {$1:1$}  (m-2-3);
    \path[stealth-stealth]        (m-2-3) edge node[above] {$1:1$}  (m-2-4);

    \path[stealth-]        (m-2-1) edge node[right] {$\subset$}  (m-1-1);
    \path[stealth-]        (m-2-2) edge node[right] {$\subset$}  (m-1-2);
    \path[stealth-]        (m-2-3) edge node[right] {$\subset$}  (m-1-3);
    \path[stealth-]        (m-2-4) edge node[right] {$\subset$}  (m-1-4);

\node at (m-1-1.north) [above=-1pt, color=black] {\smalll\sf {matrices}};
\node at (m-1-2.north) [above=-1pt, color=black] {\smalll\sf {M\"obius action}};
\node at (m-1-3.north) [above=-1pt, color=black] {\smalll\sf {reversions}};
\node at (m-1-4.north) [above=-1pt, color=black] {\smalll\sf {relativity}};

\node at (m-1-1.north) [above=7pt, color=black] {\smalll\sf {~}};
\node at (m-1-2.north) [above=7pt, color=black] {\smalll\sf {ALGEBRA}};
\node at (m-1-3.north) [above=7pt, color=black] {\smalll\sf {GEOMETRY}};
\node at (m-1-4.north) [above=7pt, color=black] {\smalll\sf {PHYSICS}};

\end{tikzpicture}   
\end{equation}
Group ${\rm SO}^\pm_{\rm o}(1,2)$ is an extended Lorentz group that besides rotations (hyperbolic and elliptic) admits also reflections in space-like directions.

\newpage
%---------------------------------------------------------------------------------------
\paragraph{C.  Beyond two dimensions -- Quaternions $\mathbb H$}
~\\

{\small
\flushright
And how the One of Time, of Space the Three, \\
Might in the Chain of Symbols girdled be.\\
---William Rowan Hamilton \\}

~\\
In order to consider more than two velocities, we need to move beyond $\mathbb C = \mathbb R^2$.  
The good news is that the above result may be extended to quaternions.  
This covers the case of $\mathbb R^{1,3}$, and thus includes the standard physical space-time.
Since quaternions do not commute, the case needs additional care.
\\ 

\noindent \textbf{Remark on quaternions and relativity theory:}  
A beginning student of science may -- upon learning about quaternions -- na\"ively hold hopes that quaternions 
have something to do with the structure of  space-time; after all the squares of the four basis elements of 
$\mathbb H$    (i.e.,  $1$, $\mathbi i$, $\mathbi j$, $\mathbi k$) have signs $(+,-, -, -)$, and that reminds the signature of the Minkowski space.  
As explained in the theory of Clifford algebras \cite{Por}, 
it turns out that quaternions represent the standard rotations of the \textit{Euclidean} spaces 
$\mathbb R^3$ and $\mathbb R^4$.  
Hence it might be a source of surprise that we shall use quaternions to represent hyperbolic composition of velocities in the Minkowski space.  
%The cost is that we must use rather the menhir representations of velocities.
\\ 

\noindent 
{\bf Basic facts:} Quaternions form a division algebra 
$\mathbb H = \rmspan \{1, \mathbi i, \mathbi j, \mathbi k\}$ with multiplication table
$$
\mathbi i^2 = \mathbi j^2= \mathbi k^2= -1
$$$$
\mathbi i\mathbi j= \mathbi k = -\mathbi j\mathbi i
$$$$
\mathbi j\mathbi k = \mathbi i = -\mathbi k\mathbi j
$$$$
\mathbi k\mathbi i = \mathbi j = -\mathbi i\mathbi k
$$
Typical element of $\mathbb H$, a quaternion, is $q = a + b\mathbi i + c\mathbi j+ d\mathbi k$.
 The conjugation is denoted by either a star or a bar and is defined 
$$
\bar{q} =  q^* = a - b\mathbi i - c\mathbi j - d\mathbi k
$$
Conjugation satisfies $(ab)^* = b^*a^*$.  
Norm squared is defined as  $\|q\|^{2} = qq^* = q^*q = a^{2} + b^{2} + c^{2} + d^{2}$  
and gives quaternions $\mathbb H$  a Euclidean structure.  
It follows that  $\|q\|^2 = \|\bar q\|^2$.  
Reciprocal (multiplicative inverse) is well-defined:   
$$
         q^{-1} = \bar{q}/\|q\|^2.
$$
%Quaternion is \textit{imaginary} (or \textit{pure}) if it belongs to $\rmIm \mathbb H = \rmspan\{\mathbi i, \mathbi j, \mathbi k\}$.  
The norm is multiplicative, that is it satisfies $|qp|  = |q | \, |p|$.  
\\

\noindent 
\textbf{Convention on quaternion fractions:} 
Quaternions are not commutative, thus there are two versions of division: left and right. We will understand quaternion \textit{fractions} 
via the right inverses and shall assume the following notation: 
$$
               \frac{p}{q} =pq^{-1} 
$$ 
Under such convention we have the following rules:
%$$
\begin{equation}
\label{eq:help2}
(i)\quad   \frac{pa}{qa} =\frac{p}{q}, \qquad      (ii)\quad   \frac{ap}{q} =a\frac{p}{q}, \qquad  (iii) \quad  \frac{p}{aq} =\frac{p}{q} a^{-1} 
\end{equation}
%$$

%---------------------------------------
\begin{proposition}  
\label{thm:qmatrix}
For any two 2-by-2 matrices with quaternion entries $M,N \in \Mat(2,\mathbb H)$ and $p,q,z \in \mathbb H$  the following holds:
%$$
\begin{equation}
\label{eq:qmatix}
\begin{aligned}
(i) &\quad (MN)z = M(Nz)\\[7pt]
(ii) &\quad  \frac{p}{q}\cdot \begin{bmatrix} a&b\\c&d\end{bmatrix} = \begin{bmatrix} pa&pb\\qc&qd\end{bmatrix}\\[7pt]
(iii)&\quad \begin{bmatrix} a&0\\0&b\end{bmatrix} \cdot z = azb^{-1}
\end{aligned}
\end{equation}
%$$
\end{proposition}

%---------------------------------------
\begin{proposition}  
\label{thm:efef}
For any two quaternions $\varepsilon, \varphi \in \mathbb H$ the following left division may be replaced by right division as follows:
%$$
\begin{equation}
\label{eq:efef}
      (1+\varphi \bar{\varepsilon })^{-1} (\varepsilon +\varphi )   =  (\varepsilon +\varphi )(1+\bar{\varepsilon }\varphi )^{-1} 
\end{equation}
%$$
\end{proposition}

\noindent
\textbf{Proof:}  
Since $|\varepsilon |^{2} =\varepsilon \bar{\varepsilon }=\bar{\varepsilon }\varepsilon $ is a real number, it commutes with any quaternion, 
thus we have $\varepsilon \bar{\varepsilon }\varphi  =  \varphi \bar{\varepsilon }\varepsilon $.  
Therefore the following is true by expansion:
$$
    (\varepsilon +\varphi )(1+\bar{\varepsilon }\varphi )   =  (1+\varphi \bar{\varepsilon })(\varepsilon +\varphi )\,,
$$
which leads directly to \eqref{eq:efef}.  \QED
\\

%\noindent
%\textbf{Proof:}  
%Since $|\varepsilon |^{2} =\varepsilon \bar{\varepsilon }=\bar{\varepsilon }\varepsilon $ is a real number, it commutes with any quaternion, 
%thus we have $\varepsilon \bar{\varepsilon }\varphi  =  \varphi \bar{\varepsilon }\varepsilon $.  
%Add to both sides $\varepsilon +\varphi +\varphi \bar{\varepsilon }\varphi $:
%$$
%\varepsilon +\varphi +\varepsilon \bar{\varepsilon }\varphi +\varphi \bar{\varepsilon }\varphi    =  \varepsilon +\varphi +\varphi \bar{\varepsilon }\varepsilon +\varphi \bar{\varepsilon }\varphi \,,
%$$ 
%Each side may be factored:
%$$
%    (\varepsilon +\varphi )(1+\bar{\varepsilon }\varphi )   =  (1+\varphi \bar{\varepsilon })(\varepsilon +\varphi )\,.
%$$
%Now by dividing on the left and on the right we obtain \eqref{eq:efef}.  \QED

\noindent This concludes our preliminary matter. Here is the main result: 

%------------------------------------
\begin{theorem}
\label{2Dematrix}  
A simple boost corresponds to quaternionic linear fractional action on $z\in S^3\subset \mathbb R^4 \cong \mathbb H$
represented by the map:
%$$
\begin{equation}
\label{eq:Hmatrixe1}
         z \ \rightarrow\  z'  = \frac{z+\varepsilon }{\bar{\varepsilon }z+1} 
                     \equiv   
                       \left[\begin{array}{cc} {1} & {\varepsilon } \\ {\bar{\varepsilon }} & {1} \end{array}\right]\cdot z
\end{equation}
%$$
($|z|  = 1$) 
where $\varepsilon$ is a quaternionic menhir for velocity $v$, 
%$$
\begin{equation}
\label{eq:Hmatrixe2}
    v = \frac{2\varepsilon }{1+\left|\varepsilon \right|^{2} } 
\end{equation}
%$$
\end{theorem}

\noindent
\textbf{Proof:}  
Any non-real quaternion $\varepsilon$ determines a complex plane $\rmspan\{1, \varepsilon\} = \mathbb C$ 
with $\rmIm \varepsilon/|\rmIm \varepsilon |$ acting as the ``$\sqrt{-1}$.  We consider two cases:
\\

\noindent \textbf{Case 1:}  
Let $\varepsilon \in \mathbb R$  be real. 
For any $z$, $z$ is either real -- and \eqref{eq:Hmatrixe1} holds by Section A, 
or it is not -- and then it reduces to the case of  complex plane $\rmspan\{\varepsilon, z\} = \mathbb R \oplus[\varepsilon]$ of Section 5B.
\\

\noindent \textbf{Case 2:}  
Assume that $\varepsilon\in\mathbb H$ is not real. 
Then the pair $\{1, \varepsilon\}$ spans a complex plane 
(with $\mathbi n = \rmIm \varepsilon/|\rmIm \varepsilon|$  playing the role of the imaginary unit). 
The geometric mutual relations between $x$, $x'$, and $\varepsilon$ is the same as between their uniformly rotated versions 
obtained by multiplying on the left by a unit quaternion $\bar{\varepsilon }/|\varepsilon |$.  Thus we have a map:
$$
\begin{aligned}
                          z &\rightarrow \bar{\varepsilon }z/|\varepsilon | \\
                         z' &\rightarrow \bar{\varepsilon }z'/|\varepsilon | \\
          \varepsilon  &\rightarrow \bar{\varepsilon }\varepsilon /  |\varepsilon |  = |\varepsilon|  \in  \mathbb R
\end{aligned}
$$
Since they lie in the same plane with $|\varepsilon|$  being real, this boils down to the case 1 above.  
We can use the above result and write the map $z \rightarrow  z'$:
$$
z  \ \rightarrow \    z'  \ =   \
          \left(\frac{\bar{\varepsilon }}{|\varepsilon |} \right)^{-1} 
              \frac{\frac{\bar{\varepsilon }z}{|\varepsilon |} +|\varepsilon |}{|\varepsilon |\frac{\bar{\varepsilon }z}{|\varepsilon |} +1}    
        \ = \  \left(\frac{\varepsilon }{|\varepsilon |} \right)\frac{\frac{\bar{\varepsilon }z}{|\varepsilon |} +|\varepsilon |}{\bar{\varepsilon }z+1} 
        \ =  \ \frac{z+\varepsilon }{\bar{\varepsilon }z+1}\,,
$$
where in the last equation we used property (\ref{eq:help2}).  
\QED

%---------------------------------
\begin{theorem}
\label{thm:Hef}
{\bf [quaternion version of menhir calculus]} 
A composition of two consecutive boosts corresponding to menhirs $\varepsilon$ and $\varphi$ in $\mathbb H \cong \mathbb R^4$ are equivalent 
to a boost corresponding to menhir
%$$
\begin{equation}
\label{eq:Hef1}
              \varepsilon \moplus \varphi \; \; =\; \; \frac{\varepsilon +\varphi }{1+\bar{\varepsilon }\varphi } 
\end{equation}
%$$
followed by rotation that in terms of quaternions is represented by 
%$$
\begin{equation}
\label{eq:Hef2}
         q \ \rightarrow \    q'   \ = \ (1+\varphi \bar{\varepsilon })\; q\; (1+\bar{\varphi }\varepsilon )^{-1} 
%$$
\end{equation}
%$$
The master equation that conveys this information may be written as alternative splits of quaternion-valued matrices:
%$$
\begin{equation}
\label{eq:Hef3}
\left[\begin{array}{cc} {1} & {\varphi } \\ {\bar{\varphi }} & {1} \end{array}\right]\; 
\left[\begin{array}{cc} {1} & {\varepsilon } \\ {\bar{\varepsilon }} & {1} \end{array}\right]  
\ =  \ 
\left[\begin{array}{cc} {1+\varphi \bar{\varepsilon }} & {0} \\ {0} & {1+\bar{\varphi }\varepsilon } \end{array}\right]\; 
\left[\begin{array}{cc} {1} & {\varepsilon \moplus \varphi } \\ {(\varepsilon \moplus \varphi)^*} & {1} \end{array}\right]
\end{equation}
%$$
where $\varepsilon \moplus \varphi $ is defined in \eqref{eq:Hef1}.
\end{theorem}

\noindent 
\textbf{Proof:} We need to be careful due to non-commutativity of quaternions.
Let us start with the composition of the two maps on the left-hand side of \eqref{eq:Hef3}:
$$
\begin{aligned}
\left[\begin{array}{cc}                1 & \varphi  \\ 
                                 \bar\varphi  & 1 \end{array}\right]
\left[\begin{array}{cc}                1 & \varepsilon  \\             
                            \bar\varepsilon   &  1 \end{array}\right]
    &= \left[\begin{array}{cc} {1+\varphi \bar{\varepsilon }} & \varepsilon + \varphi \\ 
                                           \bar\varepsilon + \bar\varphi & {1+\bar{\varphi }\varepsilon } \end{array}\right]
\\[3pt] 
 &= \left[\begin{array}{cc} {1+\varphi \bar{\varepsilon }} & 0 \\
                                         0 & {1+\bar{\varphi }\varepsilon } \end{array}\right]
       \left[\begin{array}{cc} 1 & (1+\varphi \bar{\varepsilon })^{-1}(\varepsilon + \varphi) \\ 
                                           (1+\bar{\varphi }\varepsilon )^{-1}(\bar\varepsilon + \bar\varphi) & 1 \end{array}\right]
\\[3pt] 
 &= \left[\begin{array}{cc} {1+\varphi \bar{\varepsilon }} & 0 \\
                                         0 & {1+\bar{\varphi }\varepsilon } \end{array}\right]
       \left[\begin{array}{cc} 1 & (\varepsilon + \varphi) (1+\bar\varepsilon\varphi )^{-1}\\ 
                                           ((\varepsilon + \varphi)(1+\bar\varepsilon\varphi  )^{-1})^* & 1 \end{array}\right]
%                                           (\overline{\varepsilon + \varphi)(1+\bar\varepsilon\varphi  )^{-1}} & 1 \end{array}\right]
\\[3pt]  
              &=      \left[\begin{array}{cc}  1+\varphi \bar\varepsilon & 0 \\
                                                           0 & 1+\bar\varphi\varepsilon  \end{array}\right]  \;
                         \left[\begin{array}{cc} 1 & {\tfrac{\varepsilon +\varphi }{1+\bar{\varepsilon }\varphi }}  \\ 
                                                        {\left({\tfrac{\varepsilon +\varphi }{1+\bar{\varepsilon }\varphi }} \right)^{*} } & {1} \end{array}\right]
\\[3pt]
              & =       \left[\begin{array}{cc}  1+\varphi \bar\varepsilon & 0 \\
                                                           0 & 1+\bar\varphi\varepsilon  \end{array}\right]  \;
                          \left[\begin{array}{cc} {1} & {\varepsilon \moplus \varphi } \\ 
                                                                           (\varepsilon \moplus \varphi )^* & {1} \end{array}\right]
\end{aligned}
$$
were we used  \eqref{eq:help2}.  
This establishes \eqref{eq:Hef3}.  
The second matrix on the right of \eqref{eq:Hef3}  represents a boost.  
To interpret the first as a rotation in $\mathbb H$,
 we  need $|1+\varphi \bar{\varepsilon }|\; =\; |1+\bar{\varphi }\varepsilon |$, 
which is easy to check. \QED
\\

The result may be presented in a matrix-free form:
%$$
\begin{equation}
\label{eq:bigfraction}
            z''   =   (1+\varphi \bar\varepsilon ) \; 
                      \frac{z+\frac{\varepsilon +\varphi }{1+\bar{\varepsilon }\varphi } }
                             {\left(\frac{\varepsilon +\varphi }{1+\bar{\varepsilon }\varphi } \right)^{*} z+1} \; 
                       (1+\bar\varphi \varepsilon )^{-1} \,,
\end{equation}
%$$

The last theorem encompasses all previous lower-dimensional cases. 
%Only commutativity of $\mathbb C$ or no room for rotation in $\mathbb R$ simplifies it to the previous theorems.  
We may state it as a general mathematical fact:  
Let $\mathbb  F = \mathbb  R, \mathbb C, \mathbb H$  be a division algebra.  
Special $2\times 2$ matrices are denoted: 
%$$
\begin{equation}
\label{eq:special1}
M(\varphi) = \left[\begin{array}{cc} {1} & {\varphi } \\ {\bar{\varphi }} & {1} \end{array}\right],   
\qquad 
R(\alpha,\beta) = \left[\begin{array}{cc} {\alpha } & {0} \\ {0} & {\beta } \end{array}\right]\,.
\end{equation}
%$$
for any $\alpha$, $\beta, \varphi\in \mathbb F$,  with $|\alpha| =|\beta|$. 
Then the following two factorizations are equivalent: 
$$
          M(\varphi) M(\varepsilon)   =   R(1+\varphi \bar{\varepsilon },1+\bar{\varphi }\varepsilon ) M(\varepsilon\moplus\varphi)\,.
$$
It is only a matter of interpretation that we associate with these objects the following meanings:

\begin{itemize}
\item
$\varphi$  =  ``Poincar\'e square root of velocity'':  $v = 2\varphi/(1+|\varphi|^2$),   
and M($\varphi$) = matrix of M\"obius action on the sphere of unit numbers $x \in \mathbb F:  \|x\|^2 = 1$. 
\item
$R(\alpha,\beta)$ = matrix of rotation of the unit sphere in $\mathbb F$ through the ``sandwich'' action $x\rightarrow\alpha x \beta^{-1}$.
\end{itemize}

The matrices generated a symplectic matrix group
$$
{\rm Sp}(1,1) = \left\{ \; A=\begin{bmatrix} a&b\\ \bar b &  \bar a \end{bmatrix} \ | \ a,b\in\mathbb H\ \right\}
$$   
We actually utilize scaled matrices with the property $a\bar a + b\bar b <1$. The scaling is convenient in defining 
the correspondence \eqref{eq:Hmatrixe1} but is irrelevant since ultimately we deal with the projective 
group $\hbox{\rm PSp}(1,1)$.

~

The whole situation is summarized below in \eqref{eq:bigdiagramH}.

%---------------------------------------------------------
\begin{equation}
\label{eq:bigdiagramH}
\begin{tikzpicture}[baseline=-0.8ex]
    \matrix (m) [ matrix of math nodes,
                         row sep=2.5em,
                         column sep=3.2em,
                         text height=4ex, text depth=3ex] 
 {
           \quad \dgaction{\hbox{\rm SO}_{\rm o}(1,4)}{\mathbb R^{1,4}}   \quad
          & \  % \quad \dgaction{SU(2)}{P\mathbb C^2}  \quad   
          &\quad  \dgaction{\hbox{\rm Sp}(1,1)}{\mathbb H^2}  \quad    \\
           \quad \dgaction{\hbox{\rm PSO}_{\rm o}(1,4)}{~\hbox{\rm P}\mathbb R^{1,4}_{0}\cong S^3~}   \quad 
          & \quad \dgaction{\Rev(4)}{S^3\subset \mathbb R^3 \ }   \quad   
          &\quad  \dgaction{\hbox{\rm PSp}(1,1)}{S^3\subset\mathbb H}  \quad     \\
  };
            \path[stealth-]        (m-1-1) edge node[above] {\small $2:1$}  (m-1-3);
            \path[stealth-stealth]        (m-2-1) edge node[above] {\small $1:1$} node [below] {$\cong$}  (m-2-2);
            \path[stealth-stealth]        (m-2-2) edge node[above] {\small $1:1$} node [below] {$\cong$}  (m-2-3);
%vertical
    \path[-stealth]        (m-1-1) edge node[right] { }  (m-2-1);
    \path[-stealth]        (m-1-3) edge node[right] { }  (m-2-3);

\node at (m-1-1.north) [above=-1pt, color=black] {\smalll\sf {relativity}};
\node at (m-2-2.north) [above= 3pt, color=black] {\smalll\sf {reversions}};
\node at (m-1-3.north) [above=-1pt, color=black] {\smalll\sf {M\"obius action}};

\node at (m-1-1.north) [above=7pt, color=black] {\smalll\sf {PHYSICS}};
\node at (m-2-2.north) [above=11pt, color=black] {\smalll\sf {GEOMETRY}};
\node at (m-1-3.north) [above=7pt, color=black] {\smalll\sf {ALGEBRA}};

\node at (m-2-1.south) [below=3pt, color=black] {\smalll\sf {CELESTIAL}};
\node at (m-2-2.south) [below=3pt, color=black] {\smalll\sf {CROMLECH}};
\node at (m-2-3.south) [below=3pt, color=black] {\smalll\sf {numbers}};
\node at (m-2-1.south) [below=11pt, color=black] {\smalll\sf {SPHERE}};

\end{tikzpicture}   
\end{equation}

~

\newpage

%-----------------------------------------------------------------------
\noindent 
{\bf D. Back to reality: three-dimensional space and quaternions}
\\

\noindent 
In order to describe the standard three-dimensional case, we can simply reduce the above result 
to a three-dimensional subspace of $\mathbb H$, for instance the imaginary part.  
We shall identify vectors of $\mathbb R^3$ with $\rmIm \mathbb H =\rmspan \{\mathbi i, \mathbi j, \mathbi k\}$ in the obvious way. 
It is easy to check that the transformation of type \eqref{eq:special1} 
preserves the imaginary part of the unit sphere in $\mathbb H$, i.e., the unit sphere in $\rmIm \mathbb H$: 
$$
z, \varepsilon \in  \rmIm \mathbb H \cap S^3 
\quad \Rightarrow \quad 
\frac{z+\varepsilon }{1+\bar{\varepsilon }z} \in \rmIm \mathbb H\cap S^3\, .
$$
(simple calculational verification). 
Note, that the constraint to $\rmIm \mathbb H$ implies that $\bar{\varepsilon }=-\varepsilon $, 
and therefore, in the matrix representation of the rotational part an interesting thing happens:  
$
                          \bar{\varphi }\varepsilon   =  \varphi \bar{\varepsilon }  =  -\varphi \varepsilon \,,
$. 
This simplifies the transformation \eqref{eq:bigfraction} to 
$$
            z''   =   (1-\varphi \varepsilon ) \; 
                      \frac{z+\frac{\varepsilon +\varphi }{1-\varepsilon\varphi } }
                             {\left(\frac{\varepsilon +\varphi }{1-\varepsilon\varphi } \right)^{*} z+1} \; 
                       (1-\varphi \varepsilon )^{-1} \,,
$$
The exterior part of the transformation  
$$
        z'  =  q\; z'\; q^{-1}      \qquad\hbox{where}\qquad q = 1-\varphi \varepsilon \,,
$$
may readily be recognized as the standard Hamilton's trick to represent rotations of $\mathbb R^3$ 
by quaternions via adjoint action:  $z\rightarrow qzq^{-1}$ .  
The composition of two boosts splits according to:
$$
            B_{\varphi} B_{\varepsilon} \  =  \ R(1-\varphi \varepsilon )\, B(\varepsilon\moplus\varphi)  
$$
where: 
$$
\begin{aligned}
\hbox{Axis of rotation:}         & A =  \rmIm (1-\varphi \varepsilon )  =  \rmIm \varphi \varepsilon \\
\hbox{Angle of rotation:}      &    \theta =  2 \arccos (Re(1-\varphi \varepsilon ))/|1-\varphi \varepsilon|)  
\end{aligned}
$$

~

%--------------------------------------\newpage

\noindent 
\textbf{E.  Higher dimensions --- Clifford algebra}
\\

\noindent 
Minkowski spaces beyond the standard 1+3 case are interesting for applications in physics 
(Kaluza-Klein model or string theories are examples).  
To extend our model to such cases the fields $\mathbb R$, $\mathbb C$, and $\mathbb H$ must be replaced by Clifford algebras.  
Recall that for a given Euclidean space 
$(\mathbf E, \,g)$ with metric $g$, 
the universal Clifford algebra ${\rm Cliff}(\mathbf E)$ is a  $2^{(\dim \mathbf E)}$ -- dimensional space 
that can be identified with the Grassmann algebra 
$\wedge \mathbf E$ with product that for two vectors $v, w \in\mathbf E \subset{\rm Cliff}(\mathbf E)$ is
$$
           \mathbf v\mathbf w   =   -g(\mathbf v,\mathbf w)  +  \mathbf v\wedge \mathbf w
$$
In particular $\mathbf v^2 = \mathbf v \mathbf v = -\|\mathbf v\|^2\equiv v^2$. 
(We shall denote the norm with a non-bold letters:  $v=|\mathbf v|$.) 
Define conjugation in the Clifford algebra as $(\mathbf a\mathbf b)^* = \mathbf b^*\mathbf a^*$ for arbitrary elements of ${\rm Cliff}({\bf E})$, 
and $\mathbf v*= - \mathbf v$ for $\mathbf v\in E$.  
In particular, for any orthonormal basis in \textbf{E} we have 
$$
(\mathbf e_{i_1} \mathbf e_{i_2} ...\mathbf e_{i_k} )^*  =  (-\mathbf e_{i_k} )...(-\mathbf e_{i_2} )(-\mathbf e_{i_1} )
$$ 
We shall use the same convention for ``fractions'' in ${\rm Cliff}(\mathbf E)$ as in the case of quaternions:
$$
               \frac{\mathbf p}{\mathbf q}  = \mathbf p\mathbf q^{-1}
$$
for invertible elements $\mathbf q\in {\rm Cliff}(\mathbf E)$. 
%In general, such fractions are not in \textbf{$\dot{E}$} even if $p$ and $q$ are, yet the following are true: 
Analogously to the previous models, let us define :
$$
\begin{array}{lcl}
\hbox{cromlech}   &~&     K=\{\mathbf f\in \mathbf E\;|\; |\mathbf f|^2 = 1\}\\
\hbox{menhir disk} &~&   D=\{\mathbf z\in \mathbf E\;|\; |\mathbf z|^2 <1\}.
\end{array}
$$
These elements are now understood in the context of the Clifford algebra.  
%In particular
%$$
%\mathbf {fz} = -\mathbf f\cdot \mathbf z + \mathbf f\wedge \mathbf z
%$$
%Elements of $D\subset E$ are invertible,   $\mathbf f^{-1}=-\mathbf f f^{-2}$.
%
%\noindent
%2.  $\|\mathbf z\|^2 = 1$ determines a sphere $K\cong S^{n-1}\subset\ E$.
%\\

\begin{proposition} 
1.  The following M\"obius action preserves sphere $K$
$$
\mathbf z \quad \rightarrow\quad  \mathbf z' = \left[\begin{array}{cc} {1} & \mathbf f \\ -\mathbf f & {1} \end{array}\right]\cdot \mathbf z  
                        = \frac{\mathbf z+\mathbf f}{-\mathbf f\mathbf z+1} = (\mathbf z+\mathbf f)(1- \mathbf f\mathbf z)^{-1}   \in K
$$
for any $\mathbf f\in D$ and $\mathbf z \in K$.  
\end{proposition}

Note that the denominator contains in general a bivector (a component of $\mathbf f\mathbf z$).  
Yet the ``rationalization'' of the denominator leads to the stated result. 
%The other operation is
%$$
%z    \ \rightarrow \   z' = \left[\begin{array}{cc} {b} & {0} \\ {0} & {\bar b} \end{array}\right]\cdot z 
%                                 = \frac{bz}{\bar b} \equiv  bz\bar b^{-1} 
%$$
%if $b$ is of even rank, that is belongs to the even Clifford subalgebra 
%$$
%b \in Cl_{0}(\mathbf E) = \oplus_{i=0,1,\dots } \wedge^{2i} \mathbf E  
%                         =  \mathbb R  \oplus \mathbf E \wedge \mathbf E  \oplus \wedge^{4}\mathbf E \oplus \dots .
%$$
Define two types of matrices 
$$
            M(\mathbf f) = \left[\begin{array}{cc} {1} & \mathbf f \\ -\mathbf f & {1} \end{array}\right],\qquad 
            R(b) = \left[\begin{array}{cc} {b} & {0} \\ {0} & b \end{array}\right]
$$
By similar arguments as in the previous section we have a general result.

\begin{theorem}
The composition of two boosts admits an alternative decomposition as follows:
$$
M(\mathbf e) M(\mathbf f)  =  R(1-\mathbf f\mathbf e) M(\mathbf e\moplus\mathbf f)
$$
for any $\mathbf e, \mathbf f \in D$.  In explicit terms,
$$
\left[\begin{array}{cc} {1} & {\mathbf e } \\ -\mathbf e  & {1} \end{array}\right]\; 
\left[\begin{array}{cc} {1} & \mathbf f  \\ -\mathbf e  & {1} \end{array}\right]   
=  
\left[\begin{array}{cc} 1-\mathbf f \mathbf e & {0} \\ {0} & 1-\mathbf f\mathbf e \end{array}\right]\; 
\left[\begin{array}{cc} {1} & {\mathbf e \moplus \mathbf f } \\ -\mathbf e\moplus \mathbf f & {1} \end{array}\right]
$$ 
where $\varepsilon \moplus \varphi$ is defined  
$$
\mathbf e \moplus \mathbf f  = \frac {\mathbf e+ \mathbf f}{1-\mathbf e \mathbf f } 
$$
\end{theorem}
\noindent The proofs of the above claims may be based on the fact that any two vectors $\mathbf e$ and $\mathbf f$.  
Introduce an orthonormal basis $\{\mathbf e_1,\mathbf e_2\}$ in $\hbox{span}\, \{\mathbf e,\mathbf f\}$.  
The space $\rmspan\{1, \mathbf e_1,\mathbf e_2,\mathbf e_1\mathbf e_2\}$ behaves like quaternions and   
we may thus rewrite the formulas obtained in this section, part E. 
One may also simply multiply matrices  on both sides and use the fact that 
$$
(1- \mathbf f\mathbf e)(\mathbf e+\mathbf f)=(\mathbf e+\mathbf f)(1- \mathbf e\mathbf f)
$$
because 
$(\mathbf f\mathbf e )\mathbf e =    \mathbf f(\mathbf e \mathbf e) =   (\mathbf e \mathbf e) \mathbf f  =   \mathbf e (\mathbf e  \mathbf f)$,
and similarly for $\mathbf e$.

All this establishes the isomorphism 
$$
 \hbox{\rm SO}_{\rm o}(1,n)  \cong   \hbox{gen}\{ M(\mathbf f) \;|\;  \mathbf f\in  D  \}
$$
The matrices are agreed with the $\mathbb Z_2$ grading of the Clifford algebra:
in the sense that the diagonal entries are odd elements of ${\rm Cliff}(\mathbf E)$ and the off-diagonal -- the even elements.

%\noindent
%{\bf Physics interpretation:}
%Let $v\voplus w$ represent the effective boost velocity that results as a composition of two boosts: 
%by $w$, followed by $v$.  
%In order to define its algebraic dependence on $v$ and $w$, embed the space $V$ (lab) into some subspace of a Clifford algebra Cl(E) 
%$$
%             \imath :  V \rightarrow Cl(\mathbf E)
%$$
%in a way that preserves the positive definite norm in $V$.  
%Define an algebraic addition in the Clifford algebra 
%$$
%a\moplus b= \frac{a+b}{1+\overline{a}\, b} 
%$$ 
%Now, define a map that represents vectors of velocities by ``menhirs'':
%$$
%\mu:\   V \rightarrow Cl(E): \  v \rightarrow \mu (v) = \iota (v)\moplus \iota (v)= \frac{2\iota (v)}{1+|\iota (v)|^{2} } 
%$$
%Then, as before,    
%$
%\mu (v\voplus w) = \mu (v)\moplus \mu (w)\,. 
%$
~

\noindent
{\bf Remark:}
Quite interestingly, we may model a similar action by using elements of 
$$
                 \mathbb R\oplus \mathbf E    \subset   {\rm Cliff}(\mathbf E, g)
$$
whose typical element is a formal sum of a scalar and a vector, $a=\alpha+ v$.  
with conjugation of $a$ in \textbf{$\dot{E}$}  defined as $a^* \equiv \bar a = \alpha- v$. 
This however will be explored elsewhere.
% 
%Its elements are invertible, because  $\|a\|^2 = aa^* = a^*a = \alpha^2+\|v\|^2$  
%and  $a^{-1} = a^*/(aa^*)  = (\alpha-v )/(\alpha^2+\|v\|^2)$.
%\\

%------------------------------------------------------------------------------------------------------
\section{Geometric construction of the composition of velocities}
\label{sec:mehirvelocities}

Here explore the geometric side of the ``menhir calculus'' in terms of reversions, see Section 4.  
The dimension is arbitrary but the figures are drawn for the 2-dimensional case.
Recall our notation and basic facts:
$$
\begin{aligned}
D &= \hbox{unit disk of dimension } n \\
K &= \q D = \hbox{unit sphere of dimension }  n-1
\end{aligned}
$$
Reversion through a point $p\in D$ is a map denoted by bold $\mathbf p:K\to K$.
Reversions may be composed, as in Figure \ref{fig:reversions2},
and group $\Rev(K) = \rmgen\{\mathbf p \;|\; p\in D\,\}$.

\begin{figure}[H]
\centering
\includegraphics[scale=.48]{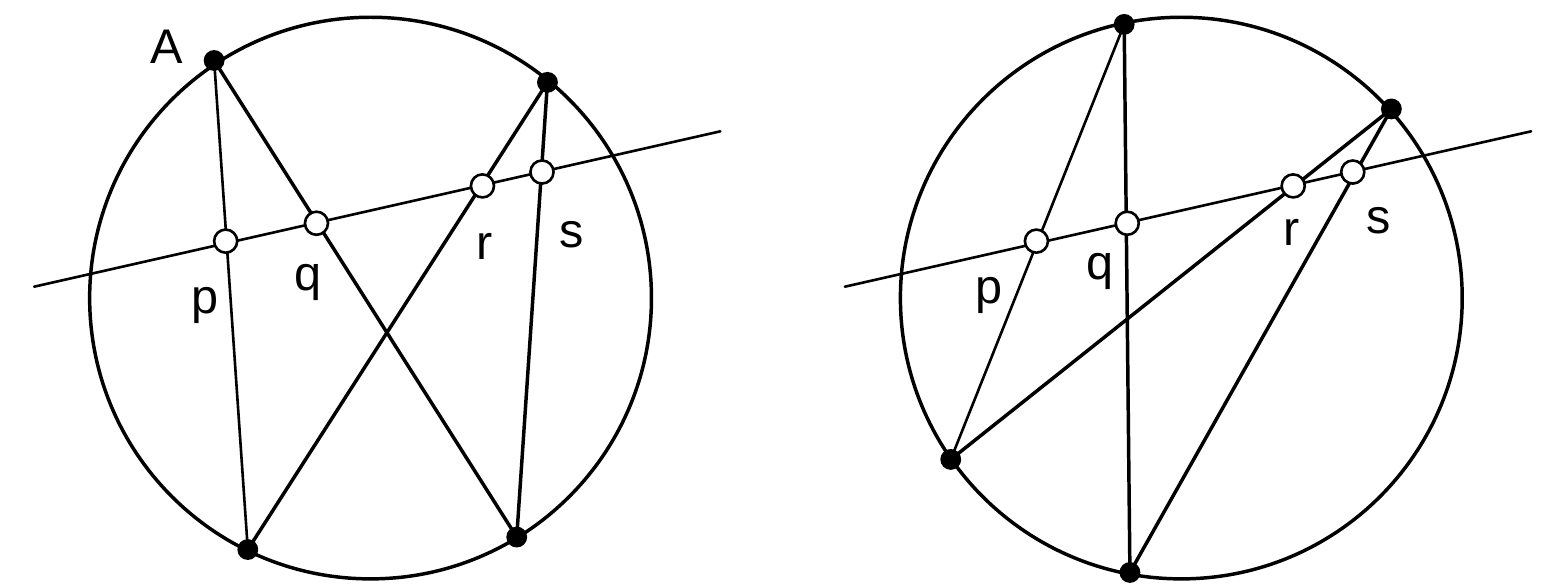}
\caption{\small The butterfly porism theorem}
\label{fig:porism}
\end{figure}

\begin{definition}
The menhir $e\in D$ of velocity $v\in D$ is defined by
$$
\mathbf e\mathbf v\mathbf e\mathbf o=\rmid \qquad \hbox{or equivalently}\qquad \mathbf o\mathbf e=\mathbf e\mathbf v.
$$
\end{definition} 
\noindent
{\bf The physical meanings:}
The unit circle $K$, cromlech,  represents horizon.  
A boost by velocity $v$  causes aberration of star positions,    
namely a star originally visible at $A\in K$ will become visible at 
$$
A' = A\mathbf o \mathbf e
$$
We start with a Lemma on butterfly, recalled here from \cite{jk-porism}:

\begin{lemma}
Suppose points $p,q,r,s\in D$  are collinear. 
Then if $A\mathbf p\mathbf q\mathbf r\mathbf s =A$ for some $A\in K$ then  $\mathbf p \mathbf q \mathbf r \mathbf s = \rmid$
(see Figure \ref{fig:porism})
\end{lemma}

%================= FIGURE 1

\begin{corollary}
For any point $a'$ on the line $(a,b)$ there exists a point $b'$ on this line such that 
$
A\mathbf a \mathbf b = A\mathbf a' \mathbf b'
$
for any $A\in K$. 
In particular 
$$
\mathbf o \mathbf e = (-\mathbf e') \mathbf o
$$
\end{corollary}

%========= Fig 
\begin{figure}[ht]
\begin{minipage}[b]{0.4\linewidth}
\centering
\includegraphics[scale=.95]{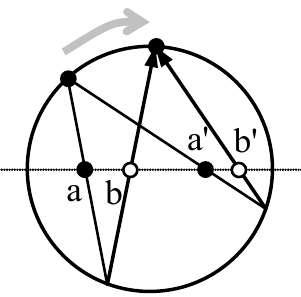}
\caption{\small $\mathbf a\mathbf b=\mathbf a'\mathbf b'$}
\label{fig:oaao}
\end{minipage}
\hspace{0.5cm}
\begin{minipage}[b]{0.5\linewidth}
      \centering
\includegraphics[scale=.95]{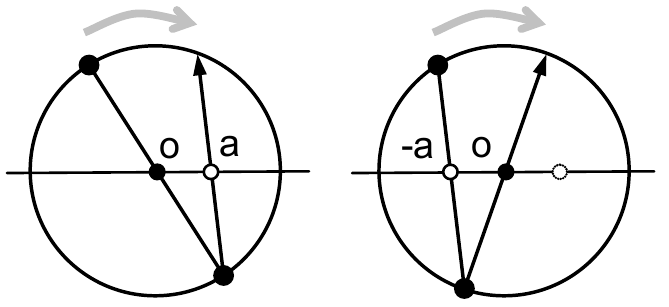}
\caption{\small $\mathbf o\mathbf a= (-\mathbf a)\mathbf o$}
\label{fig:aoa}
\end{minipage}
\end{figure}

~\\

\begin{corollary}
Composition of two boosts related to menhirs $\mathbf e$ and $\mathbf f$ may be represented by a pair of menhirs, namely
$
(-\mathbf e)\mathbf f
$.
\end{corollary}

\begin{figure}[h]
\centering
\includegraphics[scale=.9]{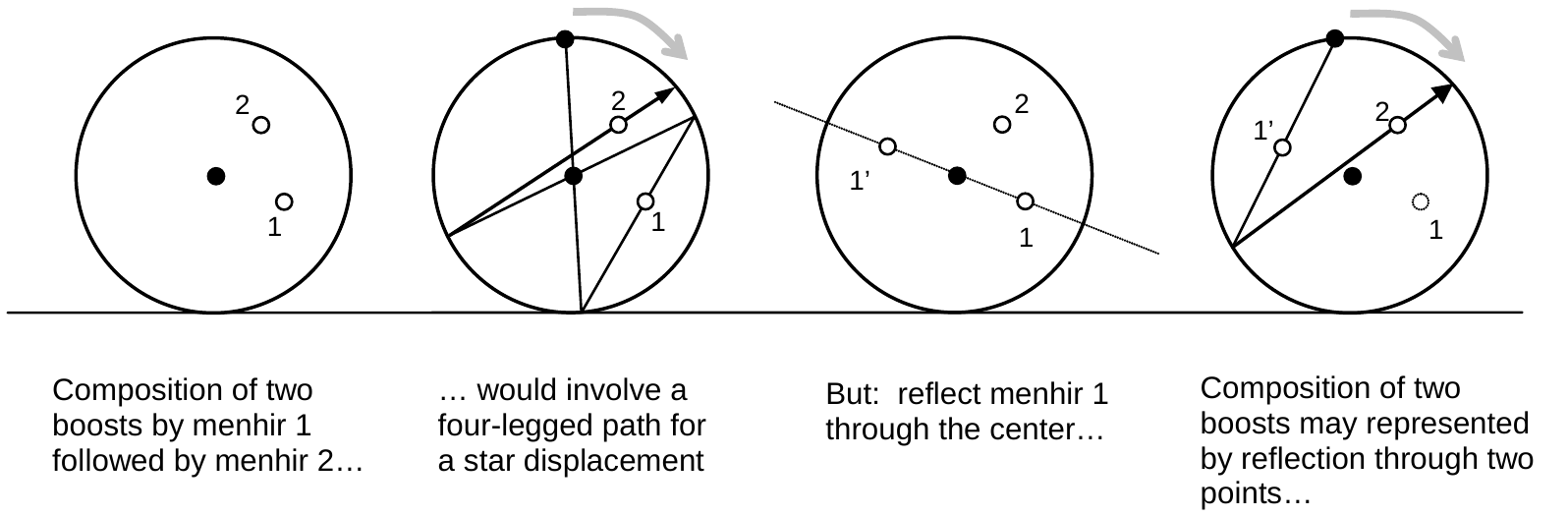}
\caption{\small Proof. Points are labeled by numbers}
\label{fig:proofadd}
\end{figure}

\noindent
{\bf Proof:}  Readily follows from the above Corollary: 
$(\mathbf o \mathbf e) (\mathbf o\mathbf f) = ((-\mathbf e) \mathbf o) (\mathbf o\mathbf f) = (-\mathbf e)\mathbf f$.  
See also Figure \ref{fig:proofadd}
\QED
\\

%========= Fig 
\begin{figure}[ht]
\begin{minipage}[b]{0.5\linewidth}
\centering
\includegraphics[scale=.89]{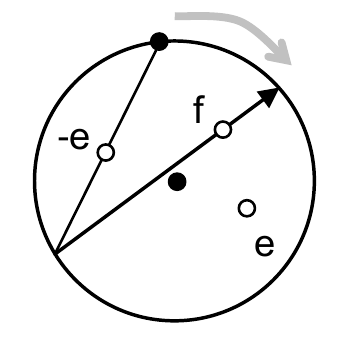}
\caption{\small Composition of two boosts in in the menhir representation}
\label{fig:reversions3}
\end{minipage}
\hspace{0.5cm}
\begin{minipage}[b]{0.5\linewidth}
      \centering
\includegraphics[scale=.79]{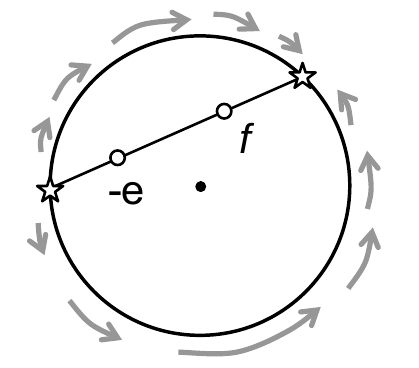}
\caption{\small Nonuniform shift of stars under a composition of two boosts}
\label{fig:shiftvw}
\end{minipage}
\end{figure}

Composition of two boosts is not a single boost.  In particular, notice that the stable points are not antipodal
(see Figure \ref{fig:shiftvw}).
But it is a composition of a single boost and a rotation.
In order to find it, we need first to ``subtract'' the rotational part.

%================= FIGURE 1

\begin{figure}[H]
\centering
\includegraphics[scale=.7]{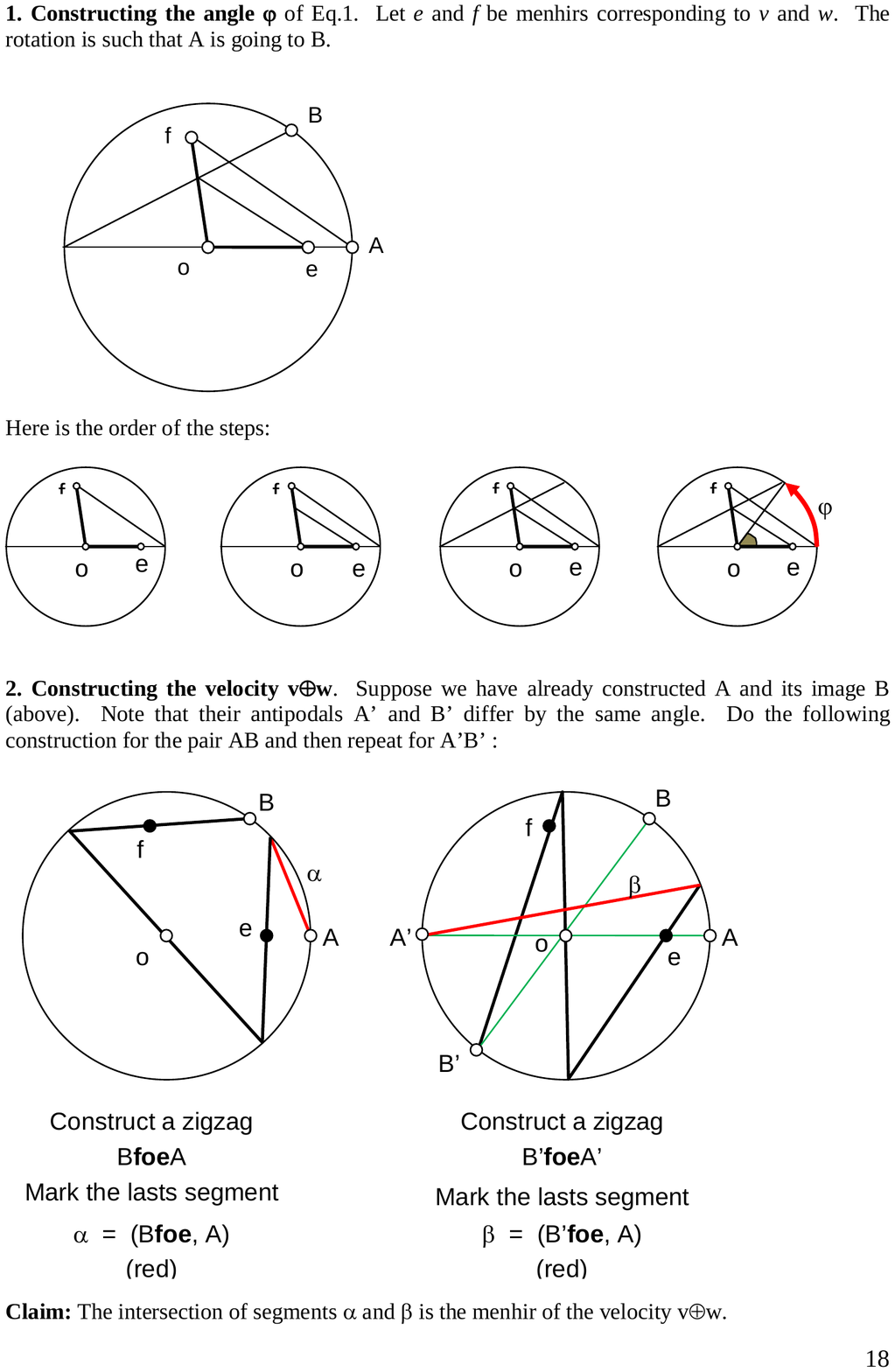}
\caption{\small The rotational component of $B(f)B(e)$ is such that $A$ goes to $B$.}
\label{fig:turn}
\end{figure}

\begin{proposition}
The rotation cosed by a composition of boosts $B(e)B(f)$  (first $e$ and then $f$)
can be constructed as shown in Figure \ref{fig:turn}.
The angle of rotation is ($AoB$).
\end{proposition}

%================= FIGURE 1

~

\noindent
Here are the steps of the construction:

%================= FIGURE 1
\begin{figure}[H]
\centering
\includegraphics[scale=.9]{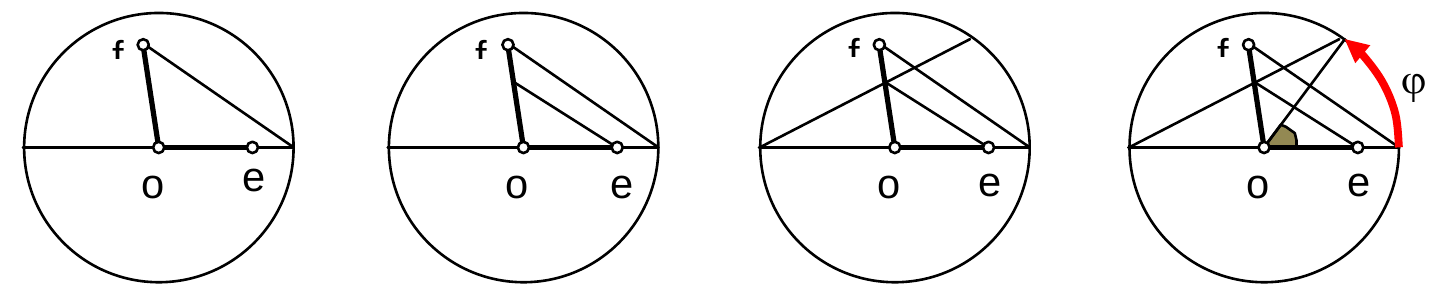}
\caption{\small Steps of the construction of the angle.}
\label{fig:rot}
\end{figure}

\noindent
{\bf Proof:} Analise Equation \ref{eq:2Dee2}.  The angle of rotation is given by 
$$
%\begin{equation}
\label{eq:policz}
\rho 
=e^{i\theta}
=  \frac{1+\varepsilon _{2} \bar{\varepsilon }_{1} }
                  {1+\bar{\varepsilon }_{2}\varepsilon _{1}  } 
=  \frac{(1+\varepsilon _{2} \bar{\varepsilon }_{1})^2 }
                  {|1+\bar{\varepsilon }_{2}\varepsilon _{1} |^2 } \,,
%\end{equation}
$$
which implies that the angle $\theta$ is twice the $\hbox{Arg}(1+\varepsilon _{2} \bar{\varepsilon }_{1})$. Hence the construction. \QED
\\

In order to construct the velocity $v\oplus w$, or rather the corresponding menhir $e\boxplus f$, 
construct first $A$ and its image $B$ as above.  
Note that their antipodals $A'$ and $B'$ also differ by the same rotation angle.  
Do the following construction for the pair $(A,B)$ and then repeat for $(A',B')$ 
to get segments $\alpha$ and $\beta$:

%================= FIGURE 1
\begin{figure}[H]
\centering
\includegraphics[scale=.7]{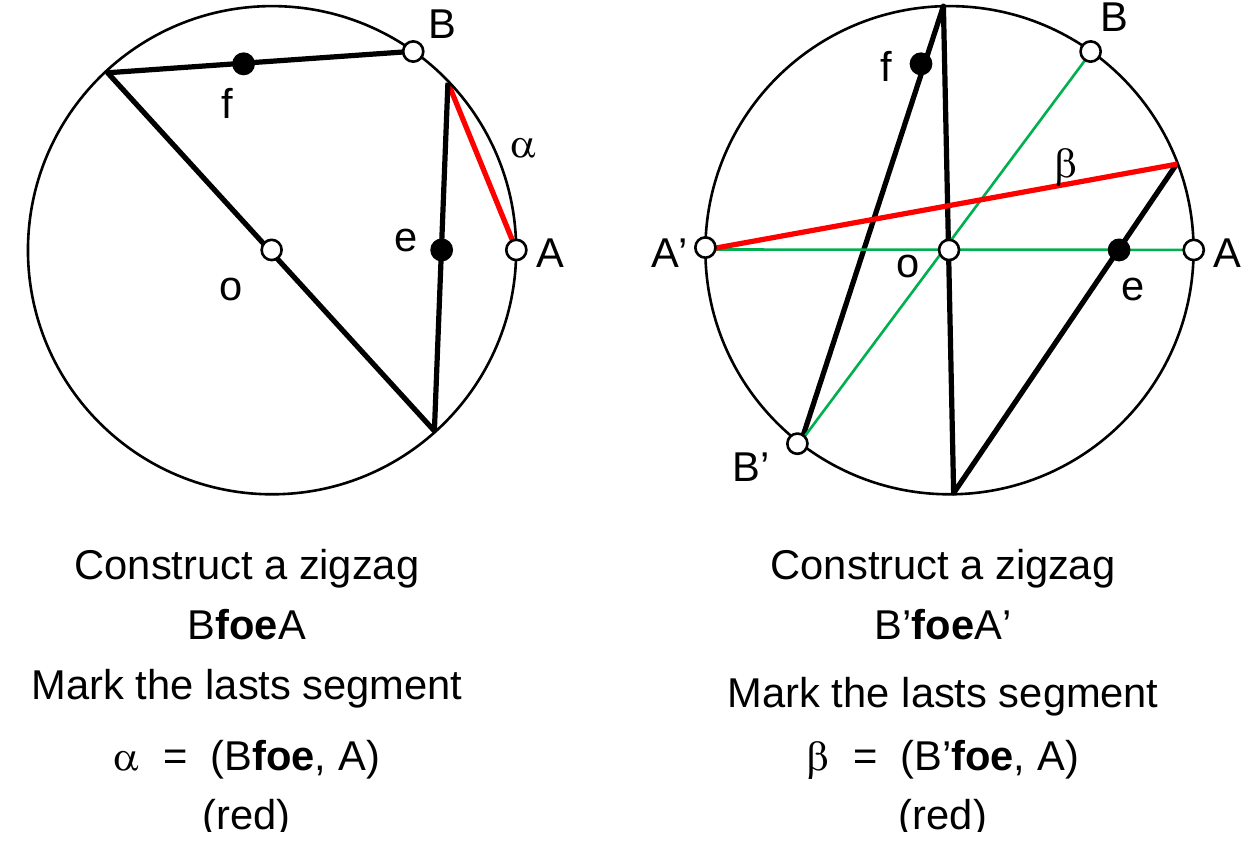}
\caption{\small $e\boxplus f = \alpha \cap \beta$.}
\label{fig:constructvw}
\end{figure}

\begin{proposition}
The intersection of segments $\alpha$ and $\beta$ is $e \boxplus f$, the menhir of the velocity $v\oplus w$.  
That is:
$$
e \boxplus f  =  (B\mathbf f\mathbf o\mathbf e ,A) \cap  (B'\mathbf f\mathbf o\mathbf  e, A)
$$
\end{proposition}

%
%\begin{corollary}
%Diagram presented in Figure \ref{fig:vdiagram} is valid.
%\end{corollary}
%\noindent
%{\bf Proof:} Exercise.

\newpage

%------------------------------------------------------------------------------------------------------
\section{Appendix}

\noindent
\textbf{\large 7.1. Dictionary for prehistoric megalithic objects}

\begin{description}
\item
\textbf{Cromlech} is a Welsh and Brytonic word for megalithic structures.  Some authors limit its usage to dolmens, but we follow the French custom to apply it to megalithic stone circles.  From \textit{cromm+llech}= \textit{bend stone}.
\item
\textbf{Menhir} is a large upright single stone that is typically a part of a larger structure system (but not a part of a single architectural construction). Breton \textit{men+hir} = \textit{stone-long}.
%\item 
%\textbf{Obelisk} is a Greek word for a tall and narrow needle-like structure made of a single or few stones. From \textit{obelos} = nail, pointed pillar.
\end{description}

\noindent
{\large\bf 7.2.  Associativity and the relativistic composition of velocities.}
\\
\\
We may now probe the feature of non-associativity of the ``addition of velocities.''
We have in general:
$$
\begin{aligned} 
(i)\quad &v \voplus (-v)  = 0\\
(ii)\quad &v \voplus w  \not= w \voplus v\\
(iii)\quad &(v\voplus w)\voplus u \ \not= \ v\voplus (w\voplus u)
\end{aligned} 
$$
This makes the the unit disk a loop.
However, the composition of the boosts understood in terms of the matrices (or reversions) {\it is} associative
(although non-commutative) and forms a group.  For instance in the quaternion case:
$$
\begin{aligned} 
(i)\quad & M_\varphi M_\varepsilon  \ \not = M_\varepsilon M_\varphi  \\
(ii)\quad & (M_\varepsilon M_\varphi) M_\delta  =M_\varepsilon (M_\varphi M_\delta)\\
\end{aligned} 
$$
where we can consider the quaternion version 
$$
M_\varepsilon
=
\begin{bmatrix} 1 &\varepsilon\\ \bar\varepsilon&1  \end{bmatrix}
$$
%To be explicit:
%$$
%\left(
%\begin{bmatrix} 1 &\varepsilon\\ \bar\varepsilon&1  \end{bmatrix}
%\begin{bmatrix} 1 &\varphi\\ \bar\varphi&1  \end{bmatrix}
%\right)\;
%\begin{bmatrix} 1 &\delta\\ \bar\delta&1  \end{bmatrix}
%\quad = \quad 
%\begin{bmatrix} 1 &\varepsilon\\ \bar\varepsilon&1  \end{bmatrix}
%\;
%\left( \begin{bmatrix} 1 &\varphi\\ \bar\varphi&1  \end{bmatrix}
%\begin{bmatrix} 1 &\delta\\ \bar\delta&1  \end{bmatrix}\right)
%$$
%But we have also 
%$$
%\begin{bmatrix} 1 &\varphi\\ \bar\varphi&1  \end{bmatrix}
%\begin{bmatrix} 1 &\varepsilon\\ \bar\varepsilon&1  \end{bmatrix}
%\quad = \quad 
%\begin{bmatrix} 1+ \varphi \bar\varepsilon&0\\ 0& 1+\varphi\bar\varepsilon  \end{bmatrix}
%\;
%\begin{bmatrix} 1 &\varphi\\ \bar\varphi&1  \end{bmatrix}
%$$
Denote 
$$
R(\varepsilon,\varphi) = \begin{bmatrix} 1+ \varphi \bar\varepsilon&0\\ 0& 1+\bar\varphi\varepsilon  \end{bmatrix}
$$
The composition of three boosts depends on the bracketing: 
%aonce we start to strip the partial result from some information, obviously the remaining becomes Same as adding with rounding to the closest integer.
%$$
%\begin{aligned}
%\left( M(a)\phantom{\Big|}M(b)\right) \,M(c)  &= M(a\moplus b) R(a,b) M(c) \\
%          &=M(a\moplus b) M(R^{-1}(a,b)c) R(a,b) \\
%          &=M((a\moplus b) \moplus R^{-1}(a,b)c)  R(a\moplus b, R^{-1}(a,b)c) ) R(a,b) \\\\
%M(a)\,\left(M(b)\phantom{\Big|} M(c)\right)  &= M(a) M(b\moplus c) R(b,c)  \\
%          &=M(a\moplus(b\moplus c))  R(a,b\moplus c) R(b,c) \\
%\end{aligned}
%$$
$$
\begin{aligned}
\left( M(a)\phantom{\Big|}M(b)\right) \,M(c)  &= M(a\moplus b) R(a,b) M(c) \\
          &=M(a\moplus b) M(R^{-1}(a,b)c) R(a,b) \\
          &=M((a\moplus b) \moplus R^{-1}(a,b)c)  R(a\moplus b, R^{-1}(a,b)c) ) R(a,b) 
\end{aligned}
$$
versus
$$
\begin{aligned}
M(a)\,\left(M(b)\phantom{\Big|} M(c)\right)  &= M(a) M(b\moplus c) R(b,c)  \\
          &=M(a\moplus(b\moplus c))  R(a,b\moplus c) R(b,c) \phantom{aaaaaaaaaaaaaaa} 
\end{aligned}
$$
The  seeming peculiarity of  ``non-associativity of relativistic addition of velocities'' 
results as a careless extension of the intuition based in the  Galilean-Newtonian physics.
%By analogy, one should not get surprised that the  addition of rational numbers like in 
%$$
%\frac{1}{2}+ \frac{2}{3} = \frac{7}{6}
%$$
%becomes ``non-associative'' when restricted to the numerators only:
%$$
%1\oplus 2 = 7\,.
%$$

%%-------------------------------------------------------------------------------------------------------------
%\newpage
%
%%---------------------------------------------------------
%\begin{equation}
%\begin{tikzpicture}[baseline=-0.8ex]
%    \matrix (m) [ matrix of math nodes,
%                         row sep=3em,
%                         column sep=4em,
%                         text height=4ex, text depth=3ex] 
% {
%             \quad  \dgaction{SO_0(1,2)}{\mathbb R^{1,2}_{0}} \quad    
%          & \quad \dgaction{\Rev_o(2)}{S^1}   \quad   
%          & \quad \dgaction{PSU(1,1)}{S^1\in\mathbb C}   \quad  \\
%%  
%\quad  \dgaction{SO^{+}(1,2)}{\mathbb R^{1,2}_{0}} \quad    
%          & \quad \dgaction{\Rev(2)}{S^1}   \quad   
%          & \quad \dgaction{PU(1,1)}{S^1\in\mathbb C}   \quad  \\
%  };
%    \path[stealth-stealth]   (m-1-1) edge node[above] {$1:1$}  (m-1-2);
%    \path[stealth-]        (m-1-2) edge node[above] {$2:1$}  (m-1-3);
%    \path[stealth-stealth]   (m-2-1) edge node[above] {$1:1$}  (m-2-2);
%    \path[stealth-]        (m-2-2) edge node[above] {$2:1$}  (m-2-3);
%
%    \path[stealth-]        (m-2-1) edge node[right] {$\subset$}  (m-1-1);
%    \path[stealth-]        (m-2-2) edge node[right] {$\subset$}  (m-1-2);
%    \path[stealth-]        (m-2-3) edge node[right] {$\subset$}  (m-1-3);
%\end{tikzpicture}   
%\end{equation}
%
%
%
%\newpage

%\newpage

~

%-------------------------------- 2 ---------------------------------------------------------------------------------------------------------------
\noindent
{\large \bf 7.2. ``Adding velocities?'' -- misunderstandings and clarifications}
\\
\\
It is sometimes stated that the essence of the theory of relativity lies in the Lorentz group.
A more accurate view seems that the heart the theory of relativity lies in geometry.
Minkowski space is a pair $(\mathbf M,g_M)$ where $\mathbf M$ is a linear space with an inner product of signature $(1,n)$
(one plus and $n$ minuses).
The Lorentz group emerges as the symmetry group of this product.
The structure of Minkowski space  -- quite like that of Euclidean space -- is entirely determined by its ``unit sphere'',  
the hyperboloid of space-like vectors 
$\|\mathbf v\|^2 = -1$, and two component hyperboloid made of time-like vectors $\|\mathbf v\|^2 = +1$ (one ``past'' and one ``future'').

One must distinguish between these three concepts: that of \textbf{velocity}, that of \textbf{observer,} and that of a \textbf{lab}.
\\

\begin{figure}[h]
\centering
\includegraphics[scale=.81]{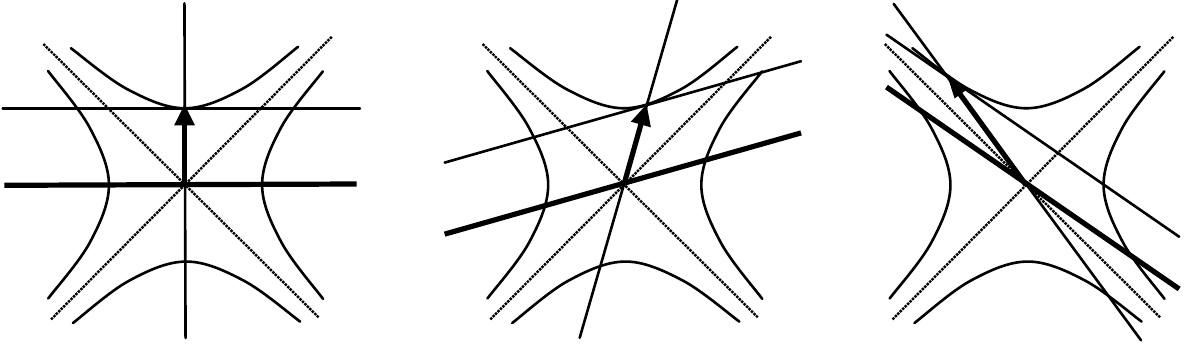}
\caption{\small Various observers come automatically with a split of space-time into ``space'' times ``time''.}
\label{fig:ST}
\end{figure}

\noindent \textbf{A.}  An \textbf{observer} at $p$ is tantamount to a unit time-like future oriented vector, \textbf{T}.  
Such a vector determines a split of the space into a Cartesian product
$$ 
               M  =  \rmspan \{\textbf{T}\} \times \textbf{T}^{\bot}   =   \hbox{\sf ``space'}' \times \hbox{\sf ``time''}
$$
of a 1-dimensional ``time axis'' span\{\textbf{T}\} and 1-codimensional 
subspace, the ``instantaneous space'' $\mathbf T^{\bot}= \{\mathbf u\in M \;|\; \mathbf u\bot \mathbf T\}$  % \{u\in M \;|\; \langle u,T\rangle\}$ 
(all vectors perpendicular to $\mathbf T$ in the sense of metric $g_M$).  
Geometrically, the space-like subspace is determined by the tangent to the sphere at $\mathbf T$, 
see Figure \ref{fig:ST} for the case $\mathbb R^{1,1}$.

%================= FIGURE 1

\noindent 
Thus the points of the upper hyperboloid parameterize observers.%
\footnote{Observer as (1,1)-tensor field in Lorentz manifold see \cite{jk-maxwell}.} 
%\footnote{The upper hyperboloid does {\it not} represent different ``velocities'' as some has suggested. 
%Velocities are pairs of points on the hyperboloid, since you need two observers to forge a velocity.}

~

\begin{figure}[h]
\centering
\includegraphics[scale=.85]{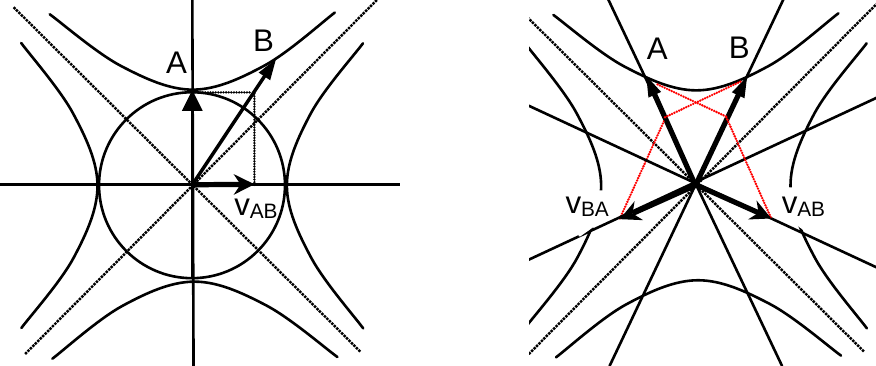}
\caption{\small Problems with misconceptualized equation ''$v +(-v) = 0$''.}
\label{fig:STvelocity}
\end{figure}

\noindent \textbf{B.}  \textbf{Velocity} is a measure of how one observer relates to another observer. 
The expression ``observer $B$ has velocity $\mathbf v_{AB}$ with respect to observer $A$'' 
should be represented as a space-like vector $\mathbf v_{AB}$ in the $A$-space, see Figure \ref{fig:STvelocity}, left.  
There is also a vector $\mathbf v_{BA}$ of the velocity of observer $A$ with respect to $B$. %, see \ref{fig:STvelocity}, center.  
What is not always clearly realized is that $\mathbf v_{AB}$ and $\mathbf v_{BA}$ are not negatives of each other, 
as one may easily see from the Figure \ref{fig:STvelocity}, right:
$$
         \mathbf v_{AB} + \mathbf v_{BA}  \neq  0 \,.
$$
Such a sum is actually a time-like past-oriented vector.  
So, how come we tend to think na\"ively that a boost followed by the ``inverse boost'' corresponds to simple expression $\mathbf v + (-\mathbf v) = 0$, obviously wrong in the light of the above?  
In order to make sense of ``addition of velocities'', we need another concept -- that of a ``lab''.
\\

%================= FIGURE

%---------------------------------
\noindent \textbf{C.}  {\bf A lab} 
represents the space as we experience it.  
We can model it as an $n$-dimensional Euclidean space $E$ and its space-time configuration as a linear map
$$
\lambda : \ \mathbf E\quad \to \quad \mathbf M \cong \mathbb R^{1,n}
$$
such that the inner product induced from the Minkowski space $\mathbf M$ agrees with that of $\mathbf E$.
Such a map defines $\mathbf T_{\lambda}$, a unit future-oriented vector perpendicular to the embedded space, 
$\mathbf T_\lambda\bot\lambda(\mathbf E)$.
The idea is that any Lorentz transformation that reorients $\mathbf E$ in $\mathbf M$ can be 
detected via pull-back map of features in $\mathbf M$ back to $\mathbf E$.

%---------------------------------------------------------
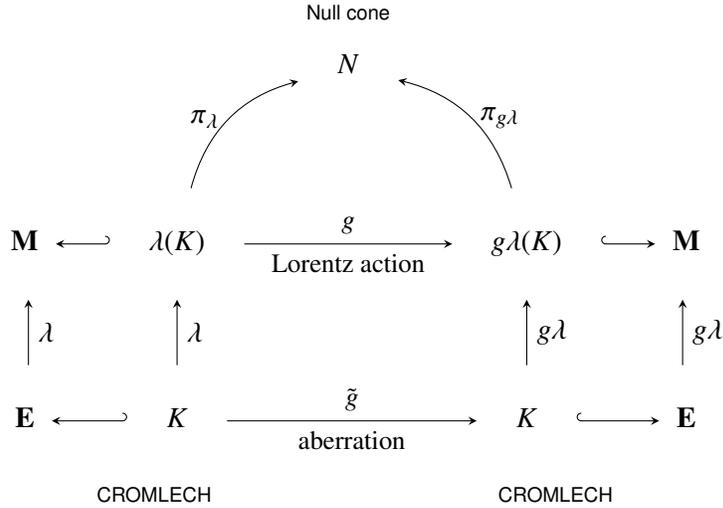
\begin{figure}[h]
\label{eq:archdiagram}
$$
\begin{tikzpicture}[baseline=-0.8ex]
    \matrix (m) [ matrix of math nodes,
                         row sep=2.3em,
                         column sep=1.9em,
                         text height=4ex, text depth=3ex] 
 {
          \ & \quad   &\quad N \quad   & \\
          \mathbf M \    
          & \quad \lambda (K)   \quad
          & \  % \quad \dgaction{SU(2)}{P\mathbb C^2}  \quad   
          &\quad g\lambda (K)\quad
          & \ \mathbf M  \\
          \mathbf E \ 
          & \quad K   \quad 
          & \quad \
          &\quad  K \quad
          &\ \mathbf E\\
  };
            \path[-stealth]        (m-2-2) edge node[above] {\small $g$} node[below] {\small Lorentz action}  (m-2-4);
            \path[-stealth]   (m-3-2) edge node[above] {\small $\tilde g$} node [below] {\small aberration}   (m-3-4);
%curved
    \path[stealth-]        (m-1-3) edge [bend right] node[left] {$\pi_\lambda$}  (m-2-2);
    \path[stealth-]        (m-1-3) edge [bend left]   node[right] {$\pi_{g\lambda}$}  (m-2-4);
    \path[-stealth]        (m-3-2) edge node[right] {$\lambda$}  (m-2-2);
    \path[-stealth]        (m-3-4) edge node[right] {$g\lambda$}  (m-2-4);
    \path[-stealth]        (m-3-1) edge node[right] {$\lambda$}  (m-2-1);
    \path[-stealth]        (m-3-5) edge node[right] {$g\lambda$}  (m-2-5);
    \path[right hook-stealth]        (m-2-4) edge node[right] {}  (m-2-5);
    \path[left hook-stealth]        (m-2-2) edge node[right] {}  (m-2-1);
    \path[right hook-stealth]        (m-3-4) edge node[right] {}  (m-3-5);
    \path[left hook-stealth]        (m-3-2) edge node[right] {}  (m-3-1);

%\node at (m-1-3.north) [above=-1pt, color=black] {\smalll\sf {M\"obius action}};
%\node at (m-2-2.north) [above=-1pt, color=black] {\smalll\sf {reversions}};
%\node at (m-1-1.north) [above=-1pt, color=black] {\smalll\sf {relativity}};

\node at (m-3-2.south) [below=0pt, color=black] {\smalll\sf {CROMLECH}\qquad ~ };
\node at (m-3-4.south) [below=0pt, color=black] {\qquad\smalll\sf {CROMLECH}};

\node at (m-1-3.north) [above=-7pt, color=black] {\smalll\sf {Null cone}};

\end{tikzpicture}   
$$
\caption{Pullback of Lorentz transformation $g$ to cromlech $K$}
\label{fig:Zbyszek}
\end{figure}

In particular, every $\lambda$ establishes a 1-1 correspondence between the celestial sphere $N={\rm P}\mathbb R^{1,n}_0$
(projective null-space) and the unit sphere $K$ in $\mathbf E$ (cromlech),
namely as the map $\lambda^{-1}\circ  \pi_\lambda$, where the projection $\pi$ is described in Section 3.
Consequently, any Lorentz map $g\in \Lambda$ of $\mathbf M$ can be pulled back to $\mathbf E$ 
as a conformal diffeomorphism $\tilde g$ of the unit sphere $K$ via the commutativity of the diagram in Figure \ref{fig:Zbyszek}.
In particular, any vector of velocity represented as $\mathbf v\in \mathbf E$ determines a hyperbolic rotation (boost) in the plane  
$\mathbf T_{\lambda}\wedge \lambda(\mathbf v)$, namely the group element   
$$
B_{\lambda,\mathbf v} =  \left(\frac{1+v}{1-v}\right)^{\frac{\mathbf T_{\lambda}\wedge \lambda(\mathbf v)}{4v}}
$$
which, from the perspective of $\mathbf E$, we simply denote as $B_{\mathbf v}$.
This pull-back of the action of the Lorentz group to the cromlech allows us 
for the geometric and algebraic constructions described in the present paper.

%A lab $(E,\lambda)$ determines an observer, but observer $\mathbf T$ defines $\lambda$
% up to the gauge group isomorphic to $S\!O(3)$.
%The metaphor of ``planetoid'' like the one in Figure \ref{fig:Stonehenge}
%represents this concept.

%
%\noindent \textbf{Remark on two addition conventions:}  
%A more conventional understanding of adding velocities is as follows:  
%Object \textit{B} moves with velocity $v$ with respect to observer \textit{A}.  
%Object \textit{C} moves with velocity w with respect to \textit{B}. 
%The sum $v\voplus w$ denotes the velocity of C as observed by \textit{A}.  
%This understanding is equivalent to our sum by planning, with order $v$ and then $w$. 
%We shall follow here the ``planetoid with a plan'' convention.  
%
%
%%================= FIGURE
%\begin{figure}[h]
%\centering
%\includegraphics[scale=.82]{C-equivalent}
%\caption{\small Two equivalent understanding of velocity addition.}
%\label{fig:equivalent}
%\end{figure}
%

%Bibliography-------------------------------------------------------------------------------

\end{document}